\documentstyle[prd,aps,preprint,epsf]{revtex}

\tighten

\newcommand{\postscript}[2]
 {\setlength{\epsfxsize}{#2\hsize}
  \centerline{\epsfbox{#1}}}

\def\mscalesym{S_m}
\def\mslopepsym{\varsigma_m^{\rm P}}
\def\mslopelsym{\varsigma_m^{\rm L}}
\def\tscalesym{S_t^{\rm SVX}}
\def\bklifesym{\tau_+^{\rm SVX}}
\def\sintbetasym{\sin2\beta}
\def\nsigsym{N_{\rm S}^{\rm SVX}}
\def\nbcksym{N_{\rm B}^{\rm SVX}}
\def\nlfracsym{F_{\rm L}^{\rm SVX}}
\def\tnegsym{\tau_-^{\rm SVX}}
\def\nfracsym{F_-^{\rm SVX}}
\def\apsym{A_{\rm P}^{\rm SST_{\rm SVX}}}
\def\epspsym{\epsilon_{\rm P}^{\rm SST_{\rm SVX}}}
\def\apctcsym{A_{\rm P}^{\rm SST_{\text{non-SVX}}}}
\def\epspctcsym{\epsilon_{\rm P}^{\rm SST_{\text{non-SVX}}}}
\def\apjchsym{A_{\rm P}^{\rm JCH}}
\def\epspjchsym{\epsilon_{\rm P}^{\rm JCH}}
\def\apselsym{A_{\rm P}^{\rm SLT}}
\def\epspselsym{\epsilon_{\rm P}^{\rm SLT}}
\def\tscaletsym{S_t^{\text{non-SVX}}}
\def\bklifetsym{\tau_+^{\text{non-SVX}}}
\def\tnegtsym{\tau_-^{\text{non-SVX}}}
\def\nlfractsym{F_{\rm L}^{\text{non-SVX}}}
\def\nfractsym{F_-^{\text{non-SVX}}}
\def\nsigctcsym{N_{\rm S}^{\text{non-SVX}}}
\def\nbckctcsym{N_{\rm B}^{\text{non-SVX}}}
\def\deltambosym{\Delta m_d}

\begin{document}

\draft
\preprint{FERMILAB-PUB-99/225-E}

\title{ \boldmath 
A measurement of $\sin 2 \beta$ from $B \to J/\psi K^0_S$
with the CDF detector}



\maketitle

\font\eightit=cmti8
\def\r#1{\ignorespaces $^{#1}$}
\hfilneg
\begin{sloppypar}
\noindent
T.~Affolder,\r {21} H.~Akimoto,\r {42}
A.~Akopian,\r {35} M.~G.~Albrow,\r {10} P.~Amaral,\r 7 S.~R.~Amendolia,\r {31} 
D.~Amidei,\r {24} J.~Antos,\r 1 
G.~Apollinari,\r {35} T.~Arisawa,\r {42}
% ...
\end{sloppypar}

\begin{center}
(CDF Collaboration)\\
\r 1  {\eightit Institute of Physics, Academia Sinica, Taipei, Taiwan 11529, 
Republic of China} \\
\r 2  {\eightit Argonne National Laboratory, Argonne, Illinois 60439} \\
\r 3  {\eightit Istituto Nazionale di Fisica Nucleare, University of Bologna,
I-40127 Bologna, Italy} \\
\r 4  {\eightit Brandeis University, Waltham, Massachusetts 02254} \\
\r 5  {\eightit University of California at Los Angeles, Los 
Angeles, California  90024} \\  
\r 6  {\eightit Instituto de Fisica de Cantabria, University of Cantabria, 
39005 Santander, Spain} \\
\r 7  {\eightit Enrico Fermi Institute, University of Chicago, Chicago, 
Illinois 60637} \\
\r 8  {\eightit Joint Institute for Nuclear Research, RU-141980 Dubna, Russia}
\\
\r 9  {\eightit Duke University, Durham, North Carolina  27708} \\
\r {10}  {\eightit Fermi National Accelerator Laboratory, Batavia, Illinois 
60510} \\
\r {11} {\eightit University of Florida, Gainesville, Florida  32611} \\
\r {12} {\eightit Laboratori Nazionali di Frascati,
         Istituto Nazionale di Fisica Nucleare, I-00044 Frascati, Italy} \\
\r {13} {\eightit University of Geneva, CH-1211 Geneva 4, Switzerland} \\
\r {14} {\eightit Harvard University, Cambridge, Massachusetts 02138} \\
\r {15} {\eightit Hiroshima University, Higashi-Hiroshima 724, Japan} \\
\r {16} {\eightit University of Illinois, Urbana, Illinois 61801} \\
\r {17} {\eightit The Johns Hopkins University, Baltimore, Maryland 21218} \\
\r {18} {\eightit Institut f\"{u}r Experimentelle Kernphysik, 
Universit\"{a}t Karlsruhe, 76128 Karlsruhe, Germany} \\
\r {19} {\eightit Korean Hadron Collider Laboratory: Kyungpook National
University, Taegu 702-701; Seoul National University, Seoul 151-742; and
SungKyunKwan University, Suwon 440-746; Korea} \\
\r {20} {\eightit High Energy Accelerator Research Organization (KEK),
         Tsukuba, Ibaraki 305, Japan} \\
\r {21} {\eightit Ernest Orlando Lawrence Berkeley National Laboratory, 
Berkeley, California 94720} \\
\r {22} {\eightit Massachusetts Institute of Technology, Cambridge,
Massachusetts  02139} \\   
\r {23} {\eightit Institute of Particle Physics: McGill University, Montreal 
H3A 2T8; and University of Toronto, Toronto M5S 1A7; Canada} \\
\r {24} {\eightit University of Michigan, Ann Arbor, Michigan 48109} \\
\r {25} {\eightit Michigan State University, East Lansing, Michigan  48824} \\
\r {26} {\eightit University of New Mexico, Albuquerque, New Mexico 87131} \\
\r {27} {\eightit The Ohio State University, Columbus, Ohio  43210} \\
\r {28} {\eightit Osaka City University, Osaka 588, Japan} \\
\r {29} {\eightit Universita di Padova, Istituto Nazionale di Fisica 
          Nucleare, Sezione di Padova, I-35131 Padova, Italy} \\
\r {30} {\eightit University of Pennsylvania, Philadelphia, 
        Pennsylvania 19104} \\   
\r {31} {\eightit Istituto Nazionale di Fisica Nucleare, University and Scuola
               Normale Superiore of Pisa, I-56100 Pisa, Italy} \\
\r {32} {\eightit University of Pittsburgh, Pittsburgh, Pennsylvania 15260} \\
\r {33} {\eightit Purdue University, West Lafayette, Indiana 47907} \\
\r {34} {\eightit University of Rochester, Rochester, New York 14627} \\
\r {35} {\eightit Rockefeller University, New York, New York 10021} \\
\r {36} {\eightit Rutgers University, Piscataway, New Jersey 08855} \\
\r {37} {\eightit Texas A\&M University, College Station, Texas 77843} \\
\r {38} {\eightit Texas Tech University, Lubbock, Texas 79409} \\
\r {39} {\eightit Istituto Nazionale di Fisica Nucleare,
                       University of Trieste/Udine, Italy} \\
\r {40} {\eightit University of Tsukuba, Tsukuba, Ibaraki 305, Japan} \\
\r {41} {\eightit Tufts University, Medford, Massachusetts 02155} \\
\r {42} {\eightit Waseda University, Tokyo 169, Japan} \\
\r {43} {\eightit University of Wisconsin, Madison, Wisconsin 53706} \\
\r {44} {\eightit Yale University, New Haven, Connecticut 06520} \\

(November 15, 1999)
\end{center}

\newpage


\begin{abstract}
This paper reports an updated  measurement of the Standard Model $CP$
violation parameter  $\sin2\beta$ using the CDF Detector at Fermilab.
The entire Run~I data sample of 110~pb$^{-1}$ of proton antiproton
collisions at $\sqrt{s}=1.8\, \rm TeV$ is used to identify a 
signal sample of 
$\sim\! 400$  $B \to J/\psi K^0_S$ events, where $J/\psi \to \mu^+\mu^-$
and $K_S^0 \to \pi^+\pi^-$. 
The flavor of the neutral $B$ meson is identified
at the time of production by combining information from
three tagging algorithms: a same-side  tag, a jet-charge tag,
and a soft-lepton tag.
A maximum likelihood fitting method is used to  determine
$\sin2\beta = 0.79 {+0.41\atop-0.44}$(stat+syst).
This value of  $\sin 2 \beta$ is consistent with the Standard Model 
prediction, based upon existing measurements,
of a large positive $CP$-violating asymmetry  in 
this decay mode.
\end{abstract}

\narrowtext

\section{Introduction}
The first observation of a violation of 
charge-conjugation parity ($CP$) invariance was 
in the neutral kaon system in 1964~\cite{cronin}.
To date, violation of $CP$ symmetry  has not
been directly observed in any other system.  
The study of $CP$ violation in the 
$B$ system is an ideal place to test the predictions of the Standard
Model~\cite{sanda,bigi,Khoze}. 
The decays of neutral $B$ mesons
into $CP$ eigenstates are of great interest,
in particular the $CP$-odd  state, $B \to J/\psi K^0_S$~\cite{rosnerisi,quinn}.
The decay $B \to J/\psi K_S^0$  
is a popular  mode in which  to observe a  $CP$-violating asymmetry
because it 
has a distinct experimental signature and 
is known theoretically to be 
free of large hadronic uncertainties~\cite{despande}. 
Furthermore, the contribution to the asymmetry 
due to penguin diagrams, which is difficult to calculate,
is negligible because the penguin contribution is small
and the tree level and penguin diagrams 
contribute with the same weak phase~\cite{pdg}. 
Previous work  searching for a $CP$-violating asymmetry in the 
decay $B \to J/\psi K^0_S$  has been presented  by  the
OPAL Collaboration~\cite{opal}.   An initial study on
the measurement of $\sin 2 \beta $ by the CDF Collaboration
is given  in Ref.~\cite{ken}.
The result reported  here incorporates and supersedes 
Ref.~\cite{ken}.
This paper reports a  measurement of $\sin 2 \beta$ that
is the best direct indication  of a $CP$-violating asymmetry  
in the neutral $B$ meson system.

\begin{figure}
\postscript{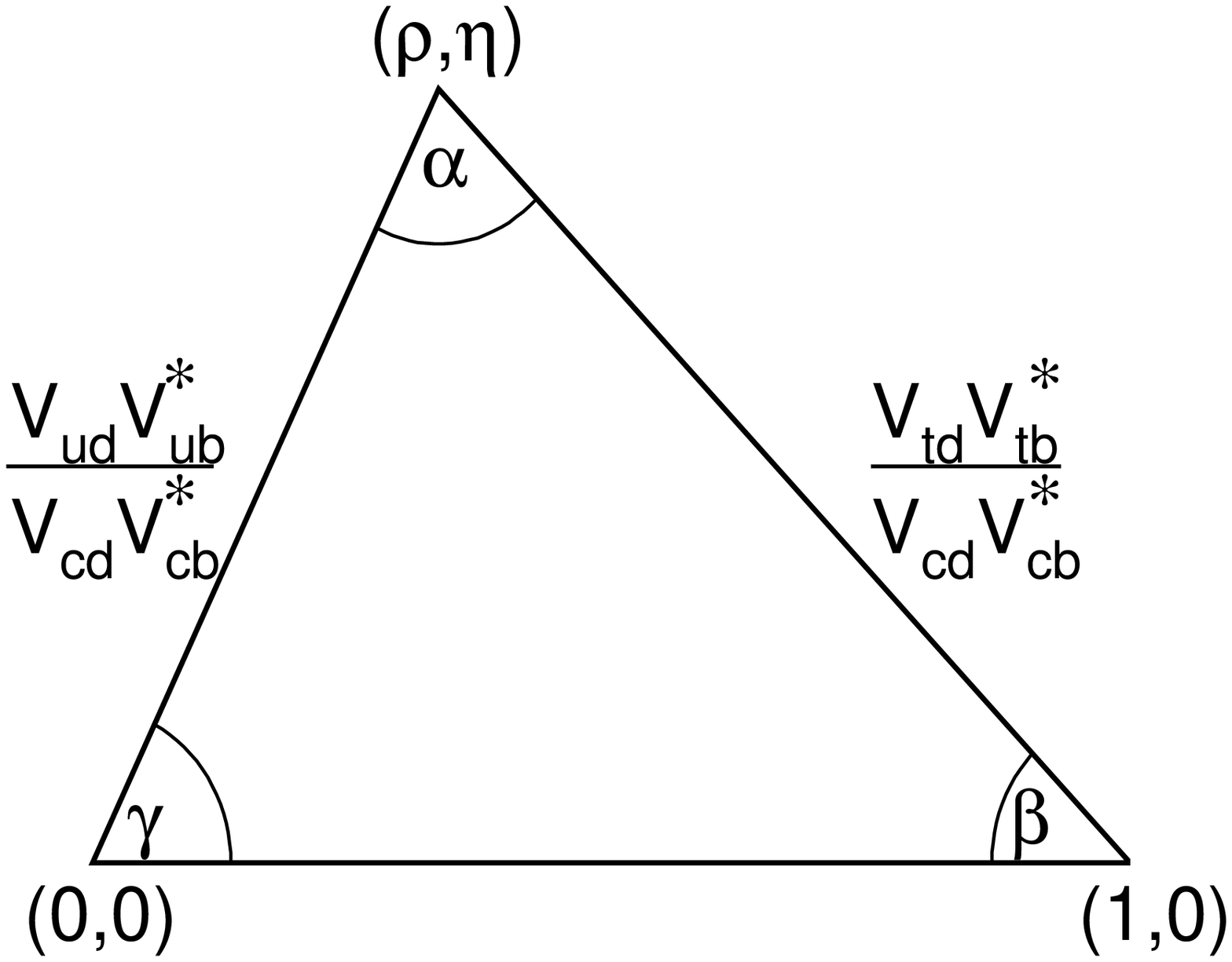}{0.750}
\caption{The unitarity triangle
indicating
the relationship between the CKM elements.}
\label{fig:triangle}
\end{figure}

Within the framework
of the Standard Model, $CP$ nonconservation  arises through a non-trivial
phase in the Cabibbo-Kobayashi-Maskawa (CKM) quark mixing matrix~\cite{ckm}.
The CKM matrix $V$ is the unitary matrix that
transforms  the mass  eigenstates into the weak
eigenstates:
\begin{eqnarray*}
  V =&&\pmatrix{V_{ud}&V_{us}&V_{ub}\cr
                V_{cd}&V_{cs}&V_{cb}\cr
                V_{td}&V_{ts}&V_{tb}\cr}\\
 \simeq && \pmatrix{1-{\lambda^2\over2}&\lambda&A\lambda^3(\rho\!-\!i\eta)\cr
                      -\lambda&1-{\lambda^2\over2}&A\lambda^2\cr
                      A\lambda^3(1\!-\!\rho\!-\!i\eta)&-A\lambda^2&1\cr}
+ O(\lambda^4).
\end{eqnarray*}
The second matrix is a useful phenomenological 
parameterization of the quark mixing matrix suggested
by Wolfenstein~\cite{lincoln}, in which $\lambda$ is the 
sine of the Cabibbo angle.
The condition of unitarity, $V^\dagger V = 1$, 
yields several relations, the most important of which is a relation between 
the first and third columns of the matrix, given by:

\[
V^*_{ub}V_{ud} 
+ V^*_{cb}V_{cd}
+ V^*_{tb}V_{td}= 0. 
\]
This relation, after division by  $V^*_{cb}V_{cd}$, 
is  displayed graphically in Fig.~\ref{fig:triangle} as a
triangle in the complex ($\rho$-$\eta$) plane, and is 
known as the unitarity triangle~\cite{linglee}.
$CP$ violation in the Standard Model 
manifests itself as  a nonzero value of $\eta$, the height of the triangle.

$CP$ nonconservation 
is expected to   manifest itself in the $B_d^0$  system
\cite{sanda} as an asymmetry in particle decay
rate versus antiparticle decay rate to a particular final state:
\[
   A_{CP} = {N(\overline{B}^0 \rightarrow J/\psi K^0_S) -
                   N(B^0 \rightarrow J/\psi K^0_S)\over {
                   N(\overline{B}^0 \rightarrow J/\psi K^0_S) +
                   N(B^0 \rightarrow J/\psi K^0_S)}} 
\]
where $N(\overline{B}^0 \rightarrow J/\psi K^0_S)$ is the number
of mesons decaying to $J/\psi K^0_S$ 
that were produced as
$\overline{B}^0$ and $N(B^0 \rightarrow J/\psi K^0_S)$ is the number
of mesons decaying to $J/\psi K^0_S$ that  were produced
as $B^0$\cite{bigi}. It should be noted that the definition
of $A_{CP}$ is the negative of that in Refs.~\cite{pdg} and \cite{opal}.

In the  Standard Model, the $CP$ asymmetry in this decay mode is 
proportional to 
$\sin 2\beta$:  $A_{CP}(t) =  \sin2\beta \sin(\Delta m_dt) $,
where $\beta$ is the angle of the unitarity triangle
shown in Fig.~\ref{fig:triangle}, $t$ is the proper decay time
of the $B^0$ meson and $\Delta m_d$ is the mass difference
between the heavy and light $B^0$ mass eigenstates. 
In a hadron collider, 
$B \overline{B}$ pairs are produced as two incoherent meson states.
Consequently,
the asymmetry can be measured  as 
either a time-dependent or time-integrated quantity. 
The time-dependent analysis is however 
statistically more powerful.
In this paper, we take advantage of
this fact and employ a sample of events that have a broad range
of time resolutions.

It is possible to combine information from 
several measurements 
to indirectly constrain the allowed range of  $\sin 2 \beta$.
Based on global fits to these measurements, it is found that the
Standard Model prefers a large positive value of $\sin 2 \beta$ and
that the fits are in good agreement
with each other~\cite{nierste,london,delphi,mele}.
One recent global fit 
finds  $\sin 2 \beta=0.75 \pm 0.09$~\cite{mele}.
However, the sign of the expected asymmetry depends on the 
sign of the product of $B_B$ and $B_K$, which are the ratios 
between the short distance contributions
to $B \overline B$ and $K \overline K$ mixing respectively
and their values in the vacuum insertion
approximation~\cite{kayser}.

To measure this asymmetry, the flavor of the $B$ meson
(whether it is a $B^0$ or a $\overline{B}^0$) 
must be identified (tagged) at the time of production. 
The effectiveness of a tagging algorithm depends on both
the efficiency for assigning  a flavor tag and the probability
that the flavor tag is correct. 
The true asymmetry is
``diluted'' by misidentifying  a $B^0$ meson as a $\overline{B}^0$ meson
or {\it vice versa}. We define the  tagging dilution  as 
$D = (N_R - N_W)/(N_R + N_W)$,
where  $N_R(N_W)  $ is the number of right  (wrong) tags.
The  observed asymmetry, given by  $A_{CP}^{\rm obs} = DA_{CP}$, is reduced
in magnitude by this dilution parameter.
As can be seen from the relation above, maximal sensitivity
to the asymmetry is achieved when the dilution factor is large.
The statistical uncertainty on $\sin2\beta$ is inversely proportional to
$\sqrt{\epsilon D^2}$, where  the efficiency 
$\epsilon$ is the fraction of events that are tagged. This
analysis combines three tagging algorithms 
in order to minimize the statistical
uncertainty of  the measurement.

\subsection{The CDF detector}

The CDF detector is described in detail elsewhere \cite{detector,TopPRD}.
The CDF detector systems that are relevant 
for this analysis are: (i)  a silicon vertex detector (SVX)~\cite{SVX},
(ii) a  time projection chamber (VTX), (iii) a  central 
tracking chamber (CTC), 
(iv) electromagnetic and hadronic calorimeters, (v) a preshower 
detector (CPR, central preradiator), 
(vi) a shower maximum detector (CES, central electron strip chamber),
and (vii) a  muon system.
The CDF coordinate system
has the $z$-axis pointing along the proton momentum, with the $x$-axis
located in the horizontal plane of the Tevatron storage ring, pointing 
radially outward, so that the $y$-axis points up.

The SVX consists of four layers of silicon axial-strip detectors
located between radii of 2.9 and 7.9 cm and extending $\pm$ 25 cm in $z$ from
the center of the detector. The geometrical acceptance of the SVX 
is $\sim\! 60$\% because the $p\overline{p}$ interactions are 
distributed with a 
Gaussian profile  along the beam axis
with a  standard deviation of $\sim\! 30$ cm,
which is large relative to the length of the detector. 
The SVX is surrounded by the VTX, which is used
to determine the $z$ coordinate of the $p \overline p$ interaction
(the primary vertex). Momenta of charged particles are measured
in three dimensions using the CTC, an 84-layer drift chamber that covers
the  pseudorapidity interval $|\eta|<$1.1, where $\eta=-\ln[\tan (\theta/2)]$,
and the angle $\theta$ is measured from the $z$-axis.
The SVX, VTX, and CTC are immersed in a 1.4~T solenoidal
magnetic field. 
The momentum transverse to the beamline ($P_T$)
of  a charged particle is determined using the SVX and CTC detectors.
The combined CTC/SVX   $P_T$ resolution is
 $\delta P_T/P_T=[(0.001\ c/{\rm GeV}\cdot P_T)^2 + 
(0.0066)^2]^{\frac{1}{2}}$. The typical uncertainty on the  $B$ meson
decay distance   is about $60\, \rm \mu m$.
The CTC also provides measurements of 
the energy loss per unit distance, $dE/dx$, of a charged
particle.

The central and endwall
calorimeters are arranged in projective towers and cover the
central region $|\eta|<$1.05. In the
central electromagnetic calorimeter, proportional chambers (CES), are embedded
near shower maximum for position measurements. The CPR is
located on the inner face of the central calorimeter and consists
of proportional chambers. 
The muon system consists of three different subsystems each containing
four layers of drift chambers.
The central muon chambers, located
behind $\sim\! 5$ absorption lengths of calorimeter, cover 85\% of
the azimuthal angle $\phi$ in the range $|\eta|<0.6$. Gaps in $\phi$
are filled in part by the central muon upgrade chambers with
total coverage in $\phi$ of 80\% and $|\eta|<0.6$. These chambers
are located behind a total of $\sim \! 8$ absorption lengths. Finally, the
central extension muon chambers provide 67\% coverage in $\phi$
for the region $0.6<|\eta|<1.0$ behind a total of $\sim \! 6$ absorption lengths.

Muons, used to reconstruct the $J/\psi$ meson and
by the soft lepton tagging algorithm (SLT),
are identified by combining a muon track segment with
a CTC track. SVX information is used when available.
Electrons, which are used by the SLT, are identified by  combining a CTC track
with information from the central calorimeters, 
the central strip chambers, $dE/dx$, and the CPR detectors.

Dimuon events are collected using a three-level trigger. The first-level
trigger system requires two charged track segments in the muon chambers.
The second level trigger requires  a CTC track, 
with $P_T$ greater than $\sim \! 2$~GeV/$c$, to match a muon chamber 
track segment.
The third level, implemented with online track reconstruction software,
requires two oppositely charged CTC tracks to match muon track
segments and a dimuon invariant mass between 2.8 and 3.4 GeV/$c^2$.
Approximately two thirds  of all $J/\psi \to \mu^+\mu^-$
events recorded enter on a dedicated $J/\psi$ trigger,
where the two reconstructed muons are from the $J/\psi$. This fraction
is consistent with expectations. 
The majority of the remaining events, referred to as ``volunteers'',
enter the sample through:
a single inclusive  muon trigger caused by one of 
the two muons from the $J/\psi$ decay, or, through 
a dimuon trigger where one of the two trigger muons
was from the $J/\psi$ and the second ``trigger muon''
is a fake muon, primarily due to punch-through.

\subsection{Overview of the analysis}

This analysis builds on the work of several previous analyses using
the various $B$ enriched data sets recorded by the CDF detector.
The $B \to J/\psi K_S^0$ decay mode is reconstructed in a manner similar to
the CDF  measurements of the branching ratio~\cite{julio,pekka} and the $B$
lifetime~\cite{hopkins}. The three tagging algorithms are then applied to
the  $B \to J/\psi K_S^0$ sample and the observed asymmetry,
given by  $A_{CP}^{\rm obs} = DA_{CP}$, is then determined. In order to
extract a value of $\sin 2 \beta$ from the observed  asymmetry,
tagging dilution parameters are required
for the three tagging algorithms. These dilution parameters 
are determined from an analysis of the calibration samples. In particular,
the same-side tagging (SST)  dilutions are
determined from a combination of results from Ref.~\cite{ken} and 
measurements on a sample of  $\sim \! 1000$ 
$B^\pm  \to J/\psi K^{\pm}$ decays.
The  jet-charge tag algorithm (JETQ) 
and soft-lepton tag algorithm (SLT)   dilutions are determined
from the $B^\pm  \to J/\psi K^{\pm}$ sample  and 
$\sim \! 40,000$ inclusive $B \to J/\psi X$ events.
The dilutions and efficiencies are then combined for each event
and a maximum likelihood fitting procedure is used to extract the 
result for $\sin 2 \beta$. 
The fit includes the possibility that
the tagging dilutions  and efficiencies have 
inherent asymmetries. In addition,
the backgrounds, divided into prompt and long-lived categories,
are also allowed to have an asymmetry. In the end, these 
possible asymmetries are found not to be significant.

Each flavor tagging  method, SST, SLT and JETQ, has 
been previously verified in a $B^0$-$\overline{B}^0$ mixing analysis.
Our previously published measurement  of $\sin 2 \beta$ used the 
$B^0 $-$\overline{B}^0$ mixing
analysis of Ref.~\cite{petar} to establish the viability  of the
SST  method~\cite{rosner}. Here we report  work
that uses the same algorithm for events where the two muons 
are contained within the SVX
detector acceptance  and uses a modified version of the  algorithm
for  events with less precise flight path
information, {\it i.e.}\ events not fully contained within  the SVX detector
acceptance. 

The two additional  tagging 
algorithms  used  are based on the 
$B^0$-$\overline{B}^0$ mixing analysis of Ref.~\cite{owen}. 
These mixing analyses  use decays of $B$ mesons with higher $P_T$
($\sim$ a factor of two higher) 
than the $B$ mesons in this analysis. This is due to the
lower trigger threshold for $J/\psi \to \mu^+\mu^-$ than for
the inclusive lepton triggers used to select the mixing analyses samples.
The SLT algorithm is similar to that 
in Ref.~\cite{owen},  except the lepton $P_T$ threshold has been lowered
 to increase the efficiency of tagging   lower $P_T$ $B$ mesons.
The JETQ  algorithm
is also  similar to the algorithm used in the mixing analysis~\cite{owen}
except the acceptance cone defining the jet has been enlarged 
and impact parameter weighting of tracks 
has been added to reduce the fraction of incorrectly tagged events.

\section{Sample Selection}

Four event samples, $B  \to J/\psi K^0_S$,
$B^\pm \to J/\psi K^{\pm}$, inclusive $B\rightarrow 
J/\psi X$ decays, and an inclusive lepton sample~\cite{petar} are used
in the determination of $\sin 2 \beta$. 
The $B$ mesons are reconstructed using the decay modes
$J/\psi \to \mu^+\mu^-$ and $K_S^0 \to \pi^+\pi^-$.
The $B  \to J/\psi K^0_S$ candidates form  the signal sample,
the $B^\pm \to J/\psi K^{\pm}$ sample is used to determine the
tagging dilutions, and the inclusive $J/\psi$ decays are used
to constrain  ratios of efficiencies. The inclusive lepton sample
was used in Refs.~\cite{ken,petar} in  the determination of the SST dilution. 

The selection criteria are largely the same as in Ref.~\cite{ken}.
The criteria for the $B  \to J/\psi K^0_S$ sample 
provide  an optimal value of the ratio $S^2/(S+N_{\rm bck})$, where
$S$ is the number of signal events and $N_{\rm bck}$ is the 
number of background
events within three standard deviations of the $B$ mass.
The square root of this ratio  enters
into the uncertainty on the measurement of   $\sin 2 \beta$.
The $J/\psi$ is identified  by  selecting two oppositely charged
muon candidates, each with  $P_T >$~1.4~GeV/$c$. Additional selection
criteria are applied to ensure good matching between the CTC
track and the muon chamber 
track segment. A $J/\psi$ candidate is
defined as a $\mu^+\mu^-$  pair within $\pm 5\sigma$ of the 
world average mass of $3.097\, {\rm GeV}/c^2$~\cite{pdg}, 
where $\sigma$ is the mass uncertainty
calculated for each event.

The $K^0_S$ candidates are found by matching pairs of oppositely
charged tracks, assumed to be pions. The $K^0_S$ candidates are
required to travel a significant distance $L_{xy}> 5\sigma_L$,
and to have  $P_T > 700$~MeV/$c$
in order to improve the signal-to-background ratio. 
The quantity $L_{xy}={\bf X}\cdot \hat P_T$ 
is the 2-D flight distance, where ${\bf X}$ is the vector
pointing from the production vertex to the decay vertex, and
$\sigma_L$ is the
measurement uncertainty on $L_{xy}$.
This flight distance 
is used to calculate the  proper decay time  $t=L_{xy}M_0/P_T$,
where $M_0$ is the world average 
$B^0$ mass of $5.2792\, {\rm GeV}/c^2$~\cite{pdg}. 
In about 15\% of the $K_S^0$ decays, SVX information is
available for one or both tracks.
When the decay vertex location in the radial direction
is found to lie beyond  the second layer of the SVX detector,
the SVX information is not used. The $J/\psi$ and
$K^0_S$ candidates are combined into a four particle fit to the
hypothesis $B \to J/\psi K^0_S$ and  the $\mu^+ \mu^-$ and
$\pi^+ \pi^-$ are constrained to the appropriate 
masses  and separate decay vertices.
The $K^0_S$ and $B$ are  constrained to point back to their
points of origin. In order to further improve the signal-to-background ratio,
$B$ candidates are accepted for $P_T(B) > 4.5$~GeV/$c$ and fit quality 
criteria are applied to the $J/\psi$ and $B$ candidates.

The data are divided into two samples, one called the SVX sample,
the other the non-SVX sample. The SVX sample requires both
muon candidates to have  at least three 
out of four possible hits that are 
well measured by  the  silicon vertex detector.
This  is the sample of $B$ candidates  with  precise decay length information
and is  similar to the sample that was 
used in the previously published CDF 
$\sin 2 \beta$ analysis.
The non-SVX sample is  the subset of events in which 
one or both muon candidates are not measured
in the silicon vertex detector.
About 30\% of the events in this sample have
one muon candidate track with high quality SVX information. 
Events of this type
lie mostly at the boundaries of the SVX detector.

\begin{figure}
\postscript{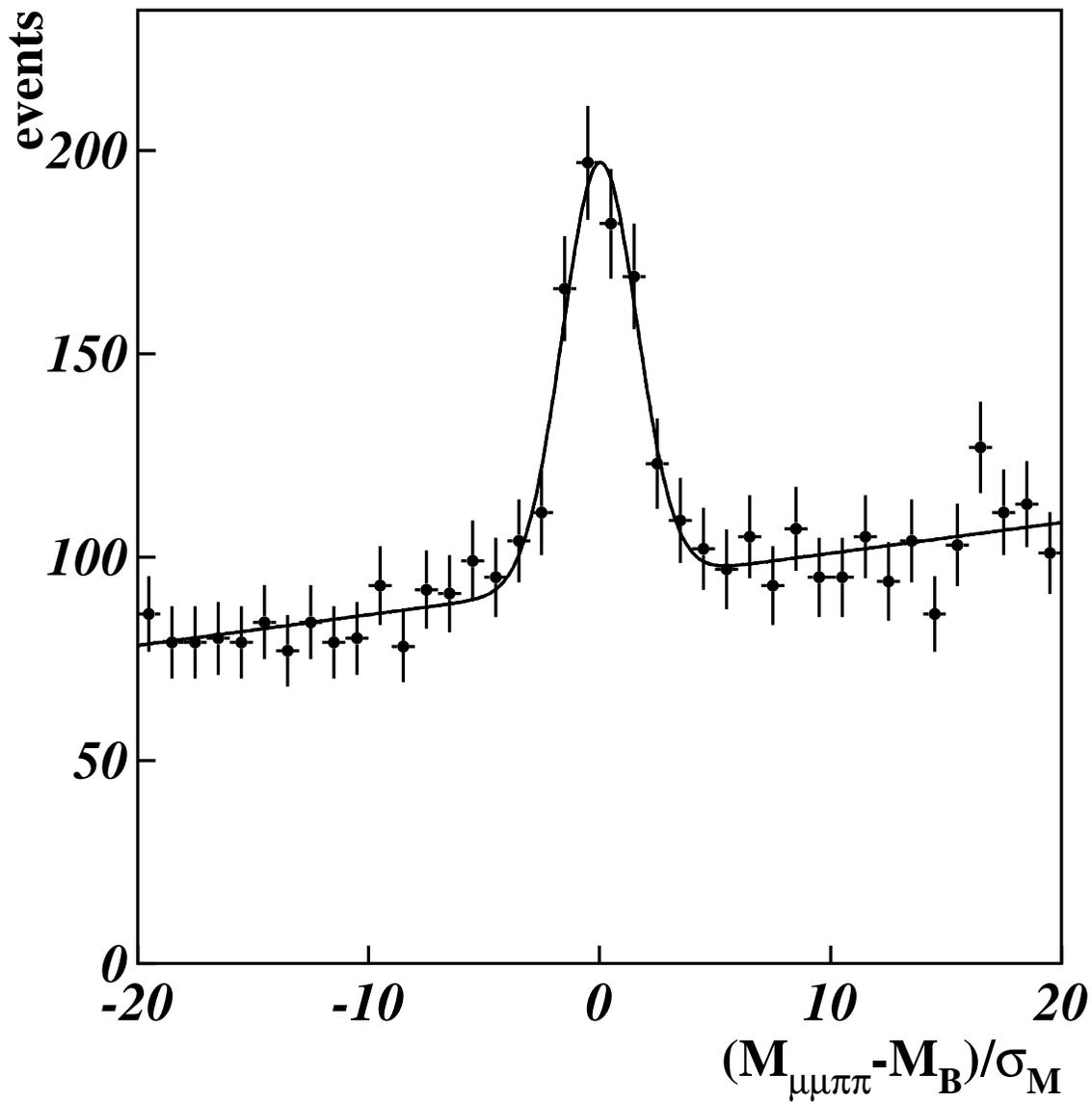}{1.0}
\caption[]{The normalized mass distribution of the $J/\psi K^0_S$
candidates. The curve is a Gaussian signal plus
linear background from a maximum  likelihood fit.}
\label{bmassall}
\end{figure}

\begin{figure}
\postscript{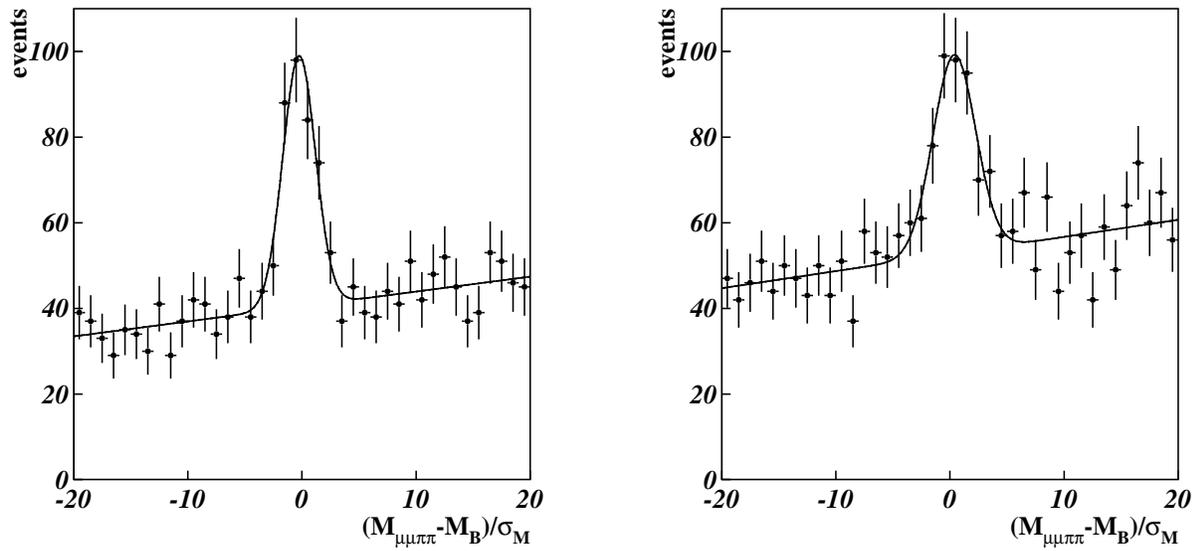}{1.0}
\caption[]{Left: \rm Normalized mass distribution of the $J/\psi K^0_S$
candidates where both muons
have good SVX information providing  a high precision
decay length measurement. 
Right: \rm 
Normalized mass distribution of the $J/\psi K^0_S$
candidates in the non-SVX sample.  
Either one or both muons
are missing good SVX information, leading to a low resolution decay length.
For both plots, the curves are Gaussian signals plus
linear background.} 
\label{bmasssvx}
\end{figure}

We define a normalized mass 
$M_N = (m_{\mu\mu\pi\pi} -M_0)/\sigma_{\rm fit}$, where 
$m_{\mu\mu\pi\pi}$  is  the four-track mass coming from the vertex
and mass-constrained fit
of the $B$ candidate. The uncertainty, $\sigma_{\rm fit}$, is from 
the fit, typically
$\sim \! 10$~MeV/$c^2$.
The normalized mass  distribution is
shown in Fig.~\ref{bmassall} and contains 4156 entries,
from which we observe  $395 \pm 31$ signal events with
a signal-to-noise ratio of 0.7.
The SVX sample  contains $202 \pm 18$ 
events (signal-to-noise ratio of 0.9)
and the non-SVX sample contains
$193\pm 26$ events (signal-to-noise ratio of 0.5)  
as shown in Fig.~\ref{bmasssvx}.  
The event yields reported here come from the full 
unbinned likelihood fit which will be described in
detail later.

\begin{figure}
\postscript{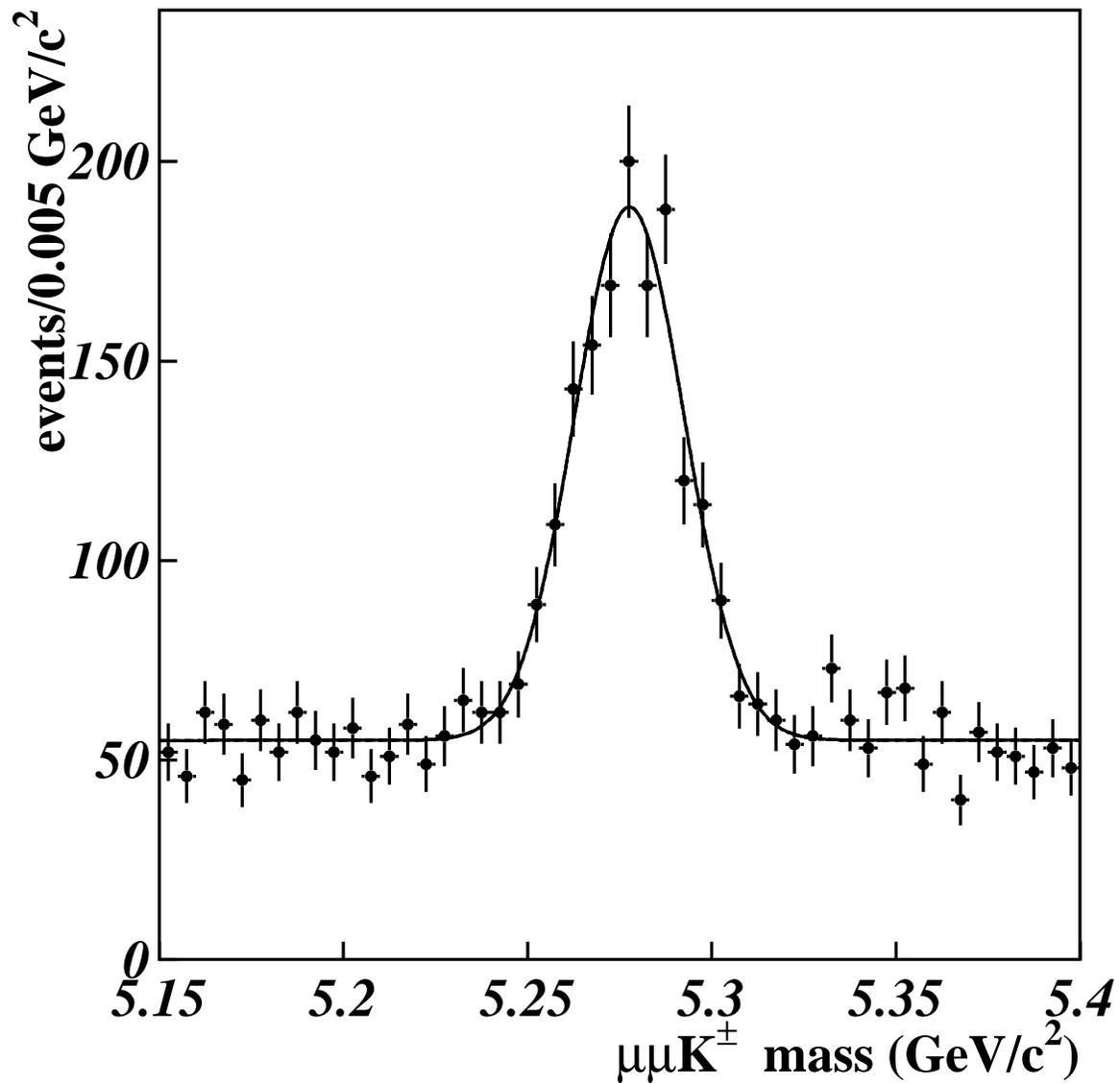}{1.0}
\caption[]{The mass distribution of the $J/\psi K^\pm$
candidates both with and without SVX information.
The curve is a Gaussian signal plus
linear background from the likelihood fit.}
\label{jpkpm}
\end{figure}

The criteria used to select the $B^\pm \to J/\psi K^{\pm}$ decays are
the same as described for $B \to J/\psi K^0_S$ decays except for the
$K^{\pm}$ selection. Since the CDF detector has limited
particle identification 
separation power at high $P_T$ using the $dE/dx$ system,
candidate kaons are defined as any track with $P_T > 2$~GeV/$c$. 
The $\mu^+\mu^-K^\pm$ mass distribution is shown in Fig.~\ref{jpkpm}
and the number of $J/\psi K^\pm$ candidates is $998\pm 51$. 

The inclusive  $J/\psi \to \mu^+\mu^-$ sample is a superset from which the
$B \to J/\psi K^0_S$ and $B^\pm \to J/\psi K^{\pm}$ samples
are derived. The inclusive sample 
is $\sim\! 80$\%  prompt $J/\psi$ from direct
$c\overline{c}$ production.
In order to enrich the sample in $B\rightarrow J/\psi X$ decays, 
both muons  are required to have good SVX information and 
the $J/\psi$  2-D 
travel distance must be $> 200$ $\mu$m  from 
the beamline. This results in a  sample of about 40,000 
$B\rightarrow J/\psi X$ decays.

\section{Tagging Algorithms}

Three tagging algorithms are used,
two opposite-side tag algorithms and one same-side  
tag (SST)  algorithm. 
The idea behind the SST algorithm~\cite{rosner} 
exploits  the local correlation between
the $B$ meson and the charge of a nearby track  to tag the flavor of 
the $B$ meson.
We employ the SST algorithm described in detail in Ref.~\cite{ken,petar}.
We consider all charged tracks that pass through all stereo 
layers of the CTC and within
a cone of radius $\Delta R=\sqrt{\Delta \eta^2 + \Delta \phi^2}<0.7$ 
centered along the $B$ meson  direction. 
Candidate tracks must
be consistent with originating from the primary vertex and have a
$P_T> 400$ MeV/$c$. If more than one candidate is found, the
track with the smallest $P_T^{\rm rel}$ is chosen, where $P_T^{\rm rel}$
is the track momentum transverse to the 
momentum sum of the track and the $B$ meson.
A tagging track with negative charge indicates a $\overline{B}^0$ meson,
while a positive  track indicates a
$B^0$ meson.

The performance of the SST algorithm could depend on the availability
of precise vertex information.  When using the SVX sample, 
the  SST algorithm of Ref.~\cite{ken} and tagging  dilution 
parameter  $D=(16.6\pm 2.2)\% $
is used. This dilution result is obtained by extrapolating
the value obtained in the mixing analysis
in Ref.~\cite{petar} to  the  lower $P_T$ of the $B \to J/\psi K_S^0$ sample.
When using the non-SVX sample, the  SST algorithm
is modified slightly by dropping  the SVX information
for all  candidate tagging tracks and adjusting the 
track selection criteria in order to increase the geometrical acceptance. 
A dilution scale factor  $f_D$, defined by
 $D_{\text{non-SVX}} = f_D D_{\rm SVX}$, is derived 
from the $B^\pm  \to J/\psi K^{\pm}$ sample.
This  relates the  SVX sample SST algorithm 
performance to that of the non-SVX sample SST algorithm. 
To measure this quantity, we compare the tagging track
using SVX information to the track we obtain when all
SVX information is ignored.  This provides a measure of 
the effectiveness of the SVX information.
We find a value of 
$f_D = (1.05 \pm 0.17)$, apply it to the 
measured SST dilution for SVX tracks, and obtain  $D=(17.4 \pm 3.6)$\%.

Opposite-side tagging refers to the identification of the flavor of the
``opposite'' $B$ in the event at the time of production.
As mentioned earlier, two algorithms are employed: soft-lepton tag (SLT)  and
jet-charge tag (JETQ) algorithms.

The SLT  algorithm is described in detail in Ref.~\cite{owen}.
The SLT algorithm associates the charge of the lepton 
(electron or muon) with the
flavor of the parent $B$-meson, which in turn is anticorrelated with  the
produced flavor of the $B$-meson that decays to $J/\psi K^0_S$.
These leptons are considered ``soft'' because their momenta
are on average considerably lower  than the high momentum
leptons from $W$ boson, $Z$ boson, and top quark  decays. 
A soft muon tag is defined as a 
charged track reconstructed
in the CTC (CTC track) with $P_T > 2$~GeV/$c$ that has been matched
to a track segment in a muon system. A soft electron tag is defined
as a CTC track with $P_T > 1$~GeV/$c$ that has been successfully
extrapolated into the calorimeters, CPR and CES detectors and passed
selection criteria. In particular,  the CPR and CES position information is
required to match with the CTC track
and the shower profiles must be consistent with an electron.
In addition, the electron candidate CTC track must have a 
$dE/dx$ deposition consistent with an electron. 
Photon  conversions
are explicitly rejected.
A dilution of  $D=(62.5\pm 14.6)\% $ is obtained
by applying the SLT algorithm to the $B^\pm \to J/\psi K^{\pm}$ sample.

If a soft lepton is not found, we try to identify a jet produced
by the opposite $B$.  We calculate a quantity called the jet
charge $Q_{jet}$ of this jet:
\[
   Q_{\rm jet} =  {{{\sum_{i} q_i{P_{T}}_i (2-(T_p)_i)}
              \over{{\sum_{i} P_{T}}_i (2-(T_p)_i)}}},
\]
where $q_i$ and ${P_{T}}_i$ are  the charge and transverse momentum
of the $i^{\rm th}$ track in the jet with $P_T > 750\, {\rm MeV}/c$.
The quantity $T_p$ is
the probability that track $i$ originated from the $p\overline{p}$
interaction
point.  
The quantity $(2-T_p)$ is constructed
such that a displaced (prompt) track has the value $T_p$ 
$\sim  \! 0$ ($1$), and the quantity $(2-T_p)$ is $\sim \! 2$ ($1$).
Tracks that arise from $B$ decays are  displaced
from the primary vertex and give a probability distribution
$T_p$ peaked near zero, lending
larger weight to the sum.   For tracks that  emanate from the
primary vertex, $T_p$ is a  flat distribution
between $0$ and $1$,  giving less weight to the jet charge 
quantity.
For $b$-quark jets, the sign of the
jet charge is on average the same as the sign of the $b$-quark that 
produced the jet, so the sign of the jet charge may be used to 
identify the flavor at production of the $B$ hadron which decayed
to $J/\psi K^0_S$.  This algorithm is conceptually similar to that
used in Ref.~\cite{owen} except that jet clustering and weighting factors
are optimized for this sample.  This optimization was performed
by  maximizing  $\epsilon D^2$ on  a sample of
$B^{\pm}\rightarrow J/\psi K^{\pm}$ events generated by a Monte Carlo 
program.

Jets are found with charged particles instead of the more commonly used 
calorimeter clusters.  The algorithm is optimized using Monte Carlo 
generated data.  All tracks in an event with $P_T > 1.75$~GeV/$c$ are 
identified as seed tracks.  For pairs of seed tracks, the quantity 
$Y_{ij}=2E_iE_j(1-\cos\theta_{ij})$ is calculated, where $E_i$, $E_j$ 
are the energies and $\theta_{ij}$ is the angle between the $i$th and 
$j$th seed tracks.  Seed tracks are combined in pairs as long as $Y_{ij}$, 
the JADE distance measure, is less than 24 GeV$^2$.  After mergings, each 
set of seed tracks defines a jet.  The remaining tracks ($P_T<1.75$~GeV/$c$) 
are combined with the jet that minimizes the distance measure provided 
that  $Y_{ij} < 24$ GeV$^2$. Any tracks unassociated with a track-group are 
discarded.   This is a modified version of the JADE clustering 
algorithm~\cite{JADE}.

Tracks within a cone of 
$\Delta R < 0.7$ with respect to the
$B \to J/\psi K_S^0$ direction are excluded from 
clustering
to avoid overlap
with the SST candidate tracks. The $B$ meson decay products 
($\mu^+$,$\mu^-$,$\pi^+$ and $\pi^-$) are 
also explicitly excluded from the track-group. 
A jet can consist of a single track with $P_T > 1.75\, {\rm GeV}/c$.
If multiple
jets are found, we choose the one 
that is most likely a $B$ jet, based on an algorithm that
uses the track impact parameter information first, if available,
and then the  jet $P_T$. 
The momentum and impact parameter weighted charge, $Q_{\rm jet}$,  
is calculated for the jet and
normalized such that $|Q_{\rm jet}| \leq 1$. 
Only tracks with $P_T > 0.750$~GeV/$c$
are used to weight the charge. The parameter $Q_{\rm jet}>0.2$ selects
the $\overline b$ quark decays and 
$Q_{\rm jet}<-0.2$ selects the $b$ quark decays. 
The value $|Q_{\rm jet}| \leq $0.2 is considered  untagged.
A dilution of $D=(23.5 \pm 6.9)\%$ is found by applying the JETQ algorithm to
the $B^\pm \to J/\psi K^{\pm}$ sample.

We use a sample of $998\pm 51$ 
$B^\pm \rightarrow J/\psi K^\pm$ decays
to determine the tagging dilutions for the opposite-side algorithms.
Using both real data and simulated data,
we have verified that 
$D(B^\pm)$ is consistent with $D(B^0)$ for the opposite-side flavor tagging
algorithms.  At the Tevatron, the strong interaction creates
$b\overline{b}$ pairs at a production energy sufficiently high 
that the fragmentation processes that create the $B$ mesons   are
largely uncorrelated. For example, the $b$ quark 
could hadronize  as a $B^-$ meson, while
independently, the $\overline{b}$ quark could hadronize
as a $B^+$, $B^0$ or $B^0_s$ meson.
These  opposite side dilution numbers are valid for both the SVX and 
non-SVX samples. The tagging dilutions and
efficiencies are presented  in Table~\ref{ta:taggers}.

Each event has the opportunity to be tagged by
two tag  algorithms: one same-side and one opposite-side.
We followed the prescription outlined in Ref.~\cite{owen} in which
the SLT tag is used if both the SLT and JETQ tags are 
available. This is done to avoid correlations between
the two opposite side tagging algorithms. 
The result of the SLT algorithm is used because the  dilution of
the SLT algorithm 
is much larger than that of the JETQ algorithm.  Given the low
efficiency for lepton tags ($6\%$) the potential overlap is small.
As mentioned earlier, tracks eligible for
the SST algorithm are excluded from the JETQ track list, thus ensuring
these two algorithms are  orthogonal.
There is however an  overlap 
between the SST and the SLT algorithms in which the lepton  is used
as the SST track. 
In order to use the dilution measured  in  Ref.~\cite{ken},
we use the identical SST algorithm on the SVX sample, and
therefore permit this overlap.
We allow leptons in the cone to account for  $b\overline{b}$ production
from the higher-order  gluon splitting process where the 
$b \to \ell X$ decay is located nearby the 
fully reconstructed $B\rightarrow J/\psi K^0_S$. This overlap
occurs in three  events in the signal region and 
the final result changes  negligibly
if these events are removed from the sample.

Based upon the tagging efficiency of each individual tagging
algorithm, we can calculate the expected fraction of events
which will tagged by two, one or zero algorithms.
We find the expected efficiency of each
combination of tags (\it e.g. \rm events tagged by both SST and SLT, 
events tagged by JETQ only, etc.) 
is consistent with estimates derived from a study of
tagging  efficiencies as applied to the $B^\pm \to J/\psi K^{\pm}$ sample.
Tag  efficiencies are higher, typically by $\sim \! 10\%$,
in the   trigger volunteer sample, except for the JETQ tagging
algorithm, in which the
efficiency increases by about 17\%.
These higher efficiencies  are due to the increased 
average charged-track multiplicity of the trigger volunteer  sample.
Thus trigger samples that do not include volunteers, as planned
for Run~II, will have lower tagging efficiencies.
It is found that
$\sim \! 80\%$ of the events in the entire
$B \to J/\psi K^0_S$ sample are tagged by at least one tagging algorithm.

\subsection{Tag sign definition}

An event is tagged if it satisfies the criteria of
any of the three tag algorithms.  For all tag algorithms, 
the flavor tag refers to whether 
the candidate $B \to J/\psi K^0_S$ was produced as
a $B^0$ or $\overline{B}^0$.
The sign of all tag algorithms follow 
the convention established by 
the same-side tag algorithm discussed in Ref.~\cite{ken}:
The positive tag ($+$ tag) is defined as the identification
of a $\overline{b}$-quark and therefore a $B^0$ meson.
The negative tag ($-$ tag) is defined as the identification
of a $b$-quark and therefore a $\overline{B}^0$ meson.
A  null tag (or tag 0) means the criteria of the tag algorithms
were not satisfied,
and the flavor of the $B$ is not identified.
A summary is provided in Table~\ref{ta:tagdefs}.

\section{Dilutions, Efficiencies and Tagging Asymmetries}

The dilutions and efficiencies described earlier need to be  generalized
in order to accommodate
possible detector asymmetries in  the analysis. 
For example, the CTC  has a small ($\sim \! 1\%$)
bias toward reconstructing more tracks of positive charge at low
transverse momentum. This small bias is due to the tilted drift cell
that is necessary to compensate for the Lorentz angle of the drift
electrons,  and a known
asymmetry in background tracks from beam pipe interactions. The formalism
for measuring and correcting for these possible tagging asymmetries
in this multi-tag analysis is provided below. 

For  $B$ mesons decaying  to a  $CP$ eigenstate,
the decay rate as a function of proper time $t$ can be written as
\[h_\pm(t) = {e^{-t/\tau}\over2\tau}(1\pm \Lambda_{CP}\sin(\Delta m_d t))\]
where $h_+(t)$ is the decay 
rate for $B$'s produced as type ``$+$'',
$h_-(t)$ is the decay rate for $B$'s produced as type ``$-$'',
and $\Lambda_{CP}=-\sin 2 \beta$ is the  asymmetry due to $CP$ violation.
Particle type ``$+$'' refers to a $B \to J/\psi K_S^0$ decay
and particle type ``$-$'' refers to a $\overline B \to J/\psi K_S^0$ decay.

To allow for an imperfect and (possibly) asymmetric tagging algorithm,
the following definitions are used.
For those $B$ mesons of (produced) type $+$, a fraction $\epsilon^+_R$ will
be actually tagged $+$, fraction $\epsilon^+_W$ will be tagged as $-$, and
fraction $\epsilon^+_0$ will not be tagged, {\it i.e.}\ tag 0. Similarly,
for those $B$ mesons of (produced) type $-$, $\epsilon^-_R$ will
be tagged $-$, fraction $\epsilon^-_W$ will be tagged as $+$, and
fraction $\epsilon^-_0$ will  be tagged as 0. Because, by definition,
$\epsilon^+_R + \epsilon^+_W + \epsilon^+_0 = 1$ and
$\epsilon^-_R + \epsilon^-_W + \epsilon^-_0 = 1$, there are 4 independent
numbers that characterize a general asymmetric tagging algorithm.

We define the efficiencies and  dilutions for the general
asymmetric tagging algorithm as $\epsilon_+ = (\epsilon^+_R+\epsilon^-_W)/2$,
$\epsilon_- = (\epsilon^-_R+\epsilon^+_W)/2$,
$\epsilon_0 = (\epsilon^+_0+\epsilon^-_0)/2$ and
\[D_+ = {\epsilon^+_R-\epsilon^-_W\over\epsilon^+_R+\epsilon^-_W},\quad 
  D_- = {\epsilon^-_R-\epsilon^+_W\over\epsilon^-_R+\epsilon^+_W},\quad 
  D_0 = {\epsilon^+_0-\epsilon^-_0\over\epsilon^+_0+\epsilon^-_0}.\]

The observed decay rate as a function of time for events tagged as $+$, $-$
or $0$ is   given by
\[
h_+(t) = {e^{-t/\tau}\over\tau}\epsilon_+(1 + \Lambda_{CP}D_+\sin(\Delta m_d
t)),
\]
\[
h_-(t) = {e^{-t/\tau}\over\tau}\epsilon_-(1 - \Lambda_{CP}D_-\sin(\Delta m_d
t)),
\]
and
\[
h_0(t) = {e^{-t/\tau}\over\tau}\epsilon_0(1 + \Lambda_{CP}D_0\sin(\Delta m_d 
t))
. \]
Note that   $\epsilon_+ + \epsilon_- + \epsilon_0 =1$ and
$\epsilon_+D_+ - \epsilon_-D_- + \epsilon_0D_0 =0$, 
so there are four  independent parameters remaining.
For example,
\[D_0 ={\epsilon_-D_- - \epsilon_+D_+ \over1-\epsilon_+ - \epsilon_-}.\]

\subsection{Combining tags in an event}
\label{combine}

Tagging information for each event is combined 
to reduce the uncertainty on the $CP$ asymmetry.
The tags are  weighted for each event by the
dilution of the individual tag algorithms. 
This procedure
must also combine the efficiencies in a similar manner.
The algorithm used to combine multiply-tagged events is as follows.
We define the tags for two tagging algorithms as $q_1$ and
$q_2$ (each taking the values $-1$, $0$, and $1$),
the individual dilutions as $D_1$ and $D_2$,
and the individual efficiencies as $\epsilon_{q_1}$ and $\epsilon_{q_2}$.
We then define the dilution-weighted tags 
${\cal D}_i = q_iD_i$, 
the product of the tag and the dilution. We
calculate the combined dilutions and efficiencies as
\[{\cal D}_{q_1q_2}=
   {{\cal D}_1+{\cal D}_2\over1+{\cal D}_1{\cal D}_2}\qquad\qquad
\epsilon_{q_1q_2}=
 \epsilon_{q_1}\epsilon_{q_2}(1+{\cal D}_1{\cal D}_2)\]
where ${\cal D}_{q_1q_2}$ is the combined dilution-weighted tag,
and $\epsilon_{q_1q_2}$ is the combined efficiency.
In this manner, 
tags in agreement as well as tags in conflict are handled 
properly:  in the cases where the charge of the two tags agree,
the effective dilution is increased; in the cases where the two
tags disagree, the effective dilution is decreased.

To help understand the expression for combined dilution 
${\cal D}$, we examine several 
limiting cases. In the case of a perfect first tagging algorithm,
$|{\cal D}_1|=1$, the combined tag always equals the value of
the perfect algorithm (${\cal D}_{q_1q_2}={\cal D}_1$), independently
of the second tagging algorithm. For the case where the first tagging algorithm
is random, $|{\cal D}_1|=0$, the combined tag always equals the value of
second algorithm (${\cal D}_{q_1q_2}={\cal D}_2$). In the case where
the result of first tagging algorithm is equal and opposite to the result
of the second tagging algorithm (${\cal D}_1={-\cal D}_2$), the
${\cal D}_{q_1q_2}=0$. This is expected when  the two tagging algorithms
have equal power but give the opposite answer.

To understand the combined efficiency  $\epsilon_{q_1q_2}$, we
consider an example.
There are  nine  possible efficiencies for 
the combined tagging algorithms, $\epsilon_{q_1q_2}$. 
The individual efficiencies for perfectly efficient  symmetric 
tagging algorithms have the
values  $\epsilon_+=\epsilon_-=0.5$ and $\epsilon_0=0$ 
($\epsilon_+ + \epsilon_- + \epsilon_0 =1$). 
In this case, five  of the nine combined  efficiencies are trivially zero.
For the case of  two perfect tagging algorithms giving the
opposite result (${\cal D}_1={-\cal D}_2$ and $|{\cal D}_1|=1$)
then the combined efficiency must be $\epsilon_{q_1q_2}=0$,
independent of the magnitude of $\epsilon_{q_1}$ and $\epsilon_{q_2}$. 
This is expected because, by definition, perfect tagging 
algorithms can not disagree.
There are only two remaining nonzero cases to examine
for the perfectly efficient 
tagging algorithm. For the case in which they 
agree, the combined efficiencies are $\epsilon_{+1,+1}=0.5$
and $\epsilon_{-1,-1}=0.5$.

\section{The Likelihood Function}

An extended  log-likelihood method is used to determine the best 
value for $\sin 2 \beta$, a free parameter in the fit.
It is helpful to refer to the
parameters collectively as a vector $\vec p$ with 65 components.
The remaining 64 parameters 
describe other features of the data (signal and background)
which need to be determined simultaneously, but have only technical importance.

The main ingredient of the likelihood function is the product
$\prod_i{\cal P}_i$ where $i$ runs over all the selected events
and ${\cal P}_i$
is the probability distribution in the measured quantities:
the normalized mass, the flight-time, and the tags ($q_1,q_2,q_3$). 
The tags, 
although discrete variables, are conceptually thought of as analogous
to continuous variables, such as the measured mass. 
The parameters $\vec p$ control the
shape of the ${\cal P}_i$. 
There is a separate set of 
parameters  for the SVX sample and the non-SVX sample
to control the shape of the components of
${\cal P}_i$. This is  especially important for
the parts of the function that specify the
distribution of the measured flight-time and  mass, 
but also the distribution of SST tags.

The form for ${\cal P}_i$ assumes that all events are of three types:
signal, prompt background, and long-lived background. Each possibility is
included in ${\cal P}_i$. Because the distributions
in mass, flight-time, and tag are 
different for the three types,
${\cal P}_i$ contains separate components ${\cal P}_{\rm S}$, 
${\cal P}_{\rm P}$, and
${\cal P}_{\rm L}$, which are the overall distributions for
signal, prompt background, and long-lived background respectively. Additional
parameters---a separate set of parameters for SVX and non-SVX---specify
the relative quantities of each event-type.
Each of the components ${\cal P}_{\rm S}$, ${\cal P}_{\rm P}$, and
${\cal P}_{\rm L}$ is expressed as the product of a time-function
($T_{\rm S}$, $T_{\rm P}$, $T_{\rm L}$), 
a mass-function ($M_{\rm S}$, $M_{\rm P}$, $M_{\rm L}$),
and a tagging-efficiency-function (${\cal E}_{\rm S}$, 
${\cal E}_{\rm P}$, ${\cal E}_{\rm L}$).

The time-function $T_{\rm S}$ 
is the probability distribution for the observed-time
given the observed tags, and therefore  has a dependence on the measured 
time
and its uncertainty, the measured tags and dilutions, and $\sin2\beta$.
The
$B^0$ lifetime $\tau$ and mixing parameter $\Delta m_d$ 
are constrained  at the world
averages: $\tau = (1.54\pm 0.04)\, \rm ps$ and
$\Delta m_d = (0.464\pm 0.018)\, \hbar \rm ps^{-1}$~\cite{pdg}. 
The $T_{\rm P}$ function is a simple Gaussian representing
the prompt $J/\psi$ background, and depends on the measured time and 
uncertainty.
There are two time-uncertainty
scale factors in $\vec p$, one for SVX events and
one for the non-SVX events, to allow for the possibility that the
measured time-uncertainties
are different from the true uncertainties by a constant factor.
The $T_{\rm L}$ 
function has positive and negative exponentials in time to represent
positive and negative long-lived background.  The positive long-lived
background arises primarily from real $B$ decays, while the negative
long-lived background is used to describe non-Gaussian tails in the
lifetime resolution.

The mass-function $M_{\rm S}$ 
is a Gaussian representing the normalized mass,
and also includes a mass-uncertainty
scale parameter. The mass-functions $M_{\rm P}$ and
$M_{\rm L}$ are 
linear in mass and normalized over the $\pm20\sigma$ mass window.

The tagging-efficiency-function ${\cal E}_{\rm S}$ gives the probability
of obtaining the observed combination of tags for a signal
event. In addition to the observed tags for the event,
it also depends on the individual tagging efficiencies and dilutions.
The prompt and long-lived background tagging-efficiency-functions, 
${\cal E}_{\rm P}$ and
${\cal E}_{\rm L}$, 
give the probability of obtaining the observed combination
of tags for prompt and long-lived background events; they depend
on individual background tagging efficiencies, but no dilutions are involved
because there is no right or wrong sign in tagging background.
For each individual tagging algorithm, the
efficiencies and the dilutions (each a component of $\vec p$)  float and
are allowed to be different for $+$ and $-$ tags and the corresponding
efficiencies and the dilutions for the tag-0 cases follow by normalization.
However, for the signal, there are constraints on the
individual tagging efficiencies and dilutions based on the available
measurements and their uncertainties.

\subsection{The likelihood function definition}

The negative log-likelihood $\ell(\vec p)$ is given by
\begin{eqnarray*}
\ell(\vec p)=&&\nsigsym+\nbcksym+\nsigctcsym+\nbckctcsym\\
&&-\sum_i\ln\left({\cal P}_i\right)+
\sum_j{1\over2}\left({f_j(\vec p)-\langle f_j\rangle\over\sigma_j}\right)^2
\end{eqnarray*}
The 4 free parameters
$\nsigsym$, $\nbcksym$, $\nsigctcsym$, and $\nbckctcsym$ refer to the
number of signal and background events in the SVX and non-SVX 
respectively.
The summation over $j$
represents a summation over all of the constraints we place on the parameters.
The constraints in general connect some function $f_j(\vec p)$
of the parameters with the corresponding
value $\langle f_j\rangle$ and uncertainty  $\sigma_j$
 determined by other measurements.

The summation over $i$ above runs over all data events that
satisfy our selection criteria; ${\cal P}_i$ is the probability
for the $i$th event, and implicitly depends on $\vec p$.
The function ${\cal P}_i$ is given by
\[
{\cal P}_i=N_{\rm S}{\cal P}_{\rm S} + 
N_{\rm B}\left[(1-F_{\rm L}){\cal P}_{\rm P} +
F_{\rm L}{\cal P}_{\rm L}\right]
\]
All events are classified as either type SVX or type non-SVX:
the $N_{\rm S}$, $N_{\rm B}$, and $F_{\rm L}$ in the expression above
are actually parameters $\nsigsym$, $\nbcksym$, and $\nlfracsym$
(the long-lived fraction of SVX background) for
SVX-type events and $\nsigctcsym$, $\nbckctcsym$, and $\nlfractsym$
for non-SVX-type events.  Although the lifetime resolution for non-SVX
events is poor relative to the SVX events, the information is used
in the likelihood function.

The functions ${\cal P}_{\rm S}$, ${\cal P}_{\rm P}$,
and ${\cal P}_{\rm L}$ are
the probabilities for the signal, prompt background, and long-lived
backgrounds. They are given by the products of time, mass, and
tagging-efficiency functions:
\[
{\cal P}_{\rm S}=T_{\rm S}M_{\rm S}{\cal E}_{\rm S} \qquad
{\cal P}_{\rm P}=T_{\rm P}M_{\rm P}{\cal E}_{\rm P} \qquad
{\cal P}_{\rm L}=T_{\rm L}M_{\rm L}{\cal E}_{\rm L}
\]

The signal time function is specified by
\[
T_{\rm S}={1\over2}{g}{*}{h}(t)
\qquad \sigma=S_t\sigma_t
\]
\[
h(t) = {e^{-t/\tau}\over\tau}\epsilon_{q_1q_2}
(1 + \Lambda_{CP}{\cal D}_{q_1q_2}\sin(\Delta mt))
\]
where  ${g}{*}{h}(t)$ represents the convolution of $h(t)$ with
a Gaussian of width $\sigma$
and depends implicitly on the values of the flight-time-uncertainty $\sigma$
and $\sin 2 \beta$.
The $S_t$ above is  $\tscalesym$ (the SVX lifetime error scale) for SVX events
and $\tscaletsym$ for non-SVX events. The $\sigma_t$ is the 
uncertainty on the
flight-time $t$ of the $B$-candidate, determined independently for each event.
The prompt  background allows the  determination of
$\tscalesym$ and $\tscaletsym$ using the global fit.
Knowledge of the individual tag
dilutions is incorporated  through the constraints.

The signal mass function is
\[
M_{\rm S}={1\over\sqrt{2\pi}\mscalesym}e^{-0.5(M_B/\mscalesym)^2}
\]
where $M_B$ is the normalized mass of the $B$-candidate and $\mscalesym$
is the $B$-mass error scale.

In an analogous fashion to ${\cal D}$, the combined signal
tagging-efficiency function ${\cal E}_{\rm S}$,
calculated by combining 3~tags as in section~\ref{combine},
depends on the 8~tagging
dilution components (as in Table~\ref{ta:sumall})
of $\vec p$ and the 8~individual $+$ and $-$ tagging-efficiency components.
The combined 
efficiency ${\cal E}_{\rm S}$ is the efficiency for obtaining the
particular combination of tags observed in the event.

The prompt background time and mass functions are
\[
T_{\rm P}={1\over2\sqrt{2\pi}\sigma}e^{-t^2/(2\sigma^2)}
\]
\[
\sigma=S_t\sigma_t \qquad S_t=\tscalesym {\rm~or~} \tscaletsym
\]
\[
M_{\rm P}=(1+\mslopepsym M_B)/(2W)
\qquad W=20
\]
where $W$ represents the normalized-mass window-size ($\pm20\sigma$),
and $\mslopepsym$ is the mass-slope of the prompt background.

The combined prompt-background tagging-efficiency function ${\cal E}_{\rm P}$
is given by the product of the individual prompt background
tagging-efficiencies: ${\cal E}_{\rm P}=\prod_k{\cal E}_{\rm P}^k$ where
$k$ runs over the tags.
The individual prompt background tagging-efficiencies are parameterized as
\[
{\cal E}_{\rm P}^k=
\left\{ \begin{array}{ll}
\epsilon_{\rm P}^k(1-A_{\rm P}^k)/2   &  q^k=-1 \\
1-\epsilon_{\rm P}^k            &  q^k=0  \\
\epsilon_{\rm P}^k(1+A_{\rm P}^k)/2   &  q^k=1
\end{array}
\right.
\]
where $q^k$ is the tag-result of the $k$th tagging algorithm, and 
$\epsilon_{\rm P}^k$
and $A_{\rm P}^k$ are components 
of $\vec p$ (specifically $\epspsym$, $\apsym$,
$\epspctcsym$, $\apctcsym$, $\epspjchsym$, $\apjchsym$,
$\epspselsym$, and $\apselsym$). The $A_{\rm P}^k$ parameters 
are the asymmetries
of the $k$th algorithm in tagging the prompt background.
The SST$_{\rm SVX}$ and
SST$_{\text{non-SVX}}$ are mutually 
exclusive---$k$ always runs over 3 tags.

The long-lived time function $T_{\rm L}$ is given by
\[
T_{\rm L}=
\left\{ \begin{array}{ll}
F_-{1\over2\tau_-}e^{t/\tau_-}        & t<0 \\
(1-F_-){1\over2\tau_+}e^{-t/\tau_+}  & t\ge0 
\end{array}
\right.
\]
where $F_-$ is one of $\nfracsym$ and $\nfractsym$,
$\tau_+$ is one of $\bklifesym$ and $\bklifetsym$, and 
$\tau_-$ is one of $\tnegsym$ and $\tnegtsym$.

The long-lived mass and tagging-efficiency functions are
\[
M_{\rm L}=(1+\mslopelsym M_B)/(2W)\qquad\qquad
{\cal E}_{\rm L}=\prod_k{\cal E}_{\rm L}^k
\]
\[
{\cal E}_{\rm L}^k=
\left\{ \begin{array}{ll}
\epsilon_{\rm L}^k(1-A_{\rm L}^k)/2   &  q^k=-1 \\
1-\epsilon_{\rm L}^k            &  q^k=0  \\
\epsilon_{\rm L}^k(1+A_{\rm L}^k)/2   &  q^k=1
\end{array}
\right.
\]
where the notation is exactly analogous to the $M_{\rm P}$ and
${\cal E}_{\rm P}$ defined above.

To further illustrate the role of constraint terms in the negative
log-likelihood function we highlight  the dilution constraints. There are
two dilution parameters, $D_+$ and $D_-$, per tagging method, the 8
parameters in $\ell(\vec p)$ representing the tagging dilutions that float
in the fit that  locates the minimum of $\ell(\vec p)$.  The
probability ${\cal P}_i$ of the $i$th $J/\psi K_S^0$ candidate depends
on these parameters through $T_{\rm S}$ and ${\cal E}_{\rm S}$.  Each tagging
method also has its own calibration information derived from other
decay modes. For example, the dilutions 
are constrained using results from the $J/\psi K^\pm$ calibration sample.
In addition,
the  $D_+$ and $D_-$ dilutions for the SST SVX sample are constrained to the
average dilution ($D_{\rm ave}= 16.6\pm 2.2$\%)
obtained after  extrapolating the mixing analysis
dilution to lower $P_T$~\cite{ken,petar}.  
The available calibration information for each tagging
method is represented in $\ell(\vec p)$ by  constraint
terms. These terms cause the function $\ell(\vec p)$ to increase as
the dilution parameters wander from the values preferred by the calibration.
When locating the minimum of $\ell(\vec p)$ we are then simultaneously
determining $\sin2\beta$ and the 8 dilution parameters, so that the
uncertainty on $\sin2\beta$ from the fit includes
contributions from all of the calibration uncertainties.

There are
similar constraint terms for the efficiency ratios for each tagging
method ($\epsilon_+/\epsilon_-$). 
The efficiency ratios $\epsilon_+/\epsilon_-$ for each tag
algorithm are constrained using the inclusive 
$B\rightarrow J/\psi X$ sample.  We fit the $J/\psi $ mass distributions
for the number of $+$ and $-$ tags.  The ratio of the number of $+$ tags
to the number of $-$ tags constrains
$\epsilon_+/\epsilon_-$. The $B\rightarrow J/\psi X$ sample is
assumed to have negligible intrinsic $CP$ asymmetry. 
In addition, the $B^0$ lifetime $\tau_{B^0}$  and mixing
parameter $\Delta m_d$ are free parameters in the fit, and there
are terms to constrain each to its world average~\cite{pdg}. 
The parameter  $\tau_{B^0}$ is constrained to $1.56 \pm 0.04$ ps and 
the parameter $\Delta m_d$ is constrained 
to $0.464 \pm 0.018 \,\hbar \rm ps^{-1}$.
Although constraining $\Delta m_d$ to the world average is the most natural
procedure, we also have the option of determining $\Delta m_d$ and
$\sin2\beta$ simultaneously from the $J/\psi K_S^0$ data by removing
the constraint on $\Delta m_d$.

The  calibration measurements are summarized in Table~\ref{ta:sumall}.
The efficiency ratios are consistent with expectations.  For SST,
the ratios are greater than unity due to a higher efficiency for
reconstructing tracks with positive charge in the CTC.

\subsection{Fits to toy Monte Carlo data}

As a check of the fitting procedure several sets
of $\sim \!1000$ toy Monte Carlo
data samples were generated, each set generated with a different
value of $\sin 2\beta$. 
The number of events, SVX/non-SVX ratio,
signal-to-background ratios, tagging efficiencies and dilutions, 
mass uncertainty and its scale factor, background lifetimes,
time uncertainties and scale factors, and other kinematic features
of the generation procedure were all tuned to be similar to the composition
of the data sample.

\begin{figure}
\postscript{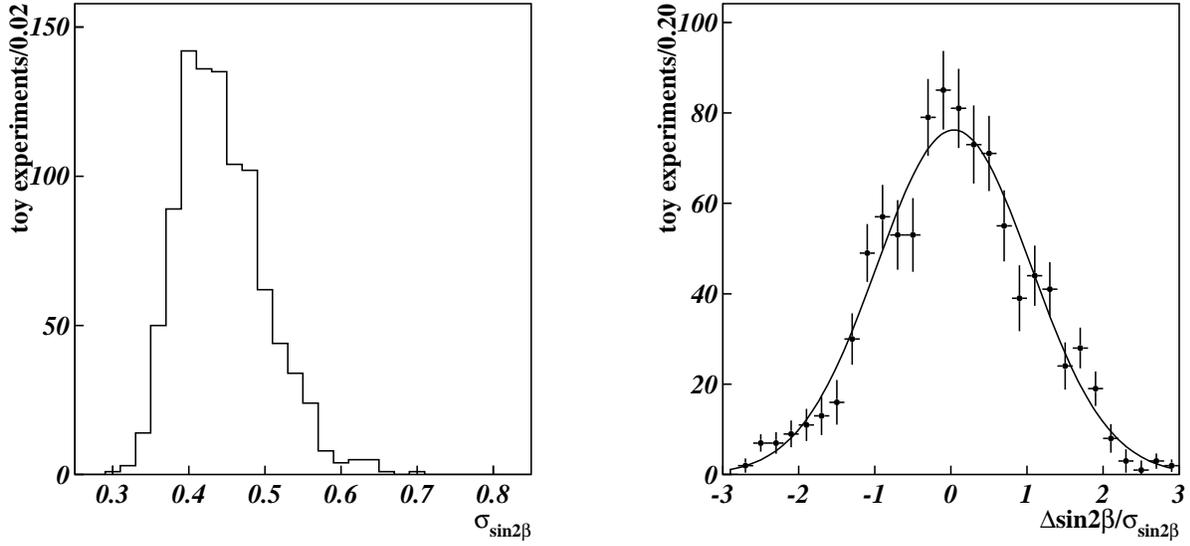}{1.0}
\caption[]{Left: Distribution of $\sigma_{\sin2\beta}$ from 
fits to multiple Monte Carlo datasets 
generated with $\sin2\beta=0.5$.
Right: Distribution of normalized $\sin2\beta$
deviations, {\it i.e.}
(fit-$\sin2\beta$ $-$ 0.5)/$\sigma_{\sin2\beta}$,
and a Gaussian fit to that distribution.  The mean
of the Gaussian fit is $0.038\pm 0.033$ and the 
width is $1.01\pm 0.03$, consistent with expectation.}
\label{multi_mc_stats}
\end{figure}

The left plot in Fig.~\ref{multi_mc_stats} shows the distribution of the
appropriate
uncertainty (allowing for asymmetric errors~\cite{MINUIT}) 
on $\sin2\beta$ returned from the Monte Carlo fits with
generated $\sin2\beta=0.5$. The
typical value of the uncertainty on $\sin2\beta$ returned from these fits is
$\sim \! 0.44$, though there is a long tail extending out to $\sim\! 0.7$.
The width of the distribution is determined by Poisson fluctuations
in the number of Monte Carlo events that are tagged.
The right plot in Fig.~\ref{multi_mc_stats} shows $[\sin2\beta({\rm
fit})-0.5]/\sigma$, where $\sigma$ is the appropriate
$+$ or $-$ uncertainty on $\sin2\beta$.

The results from this and  other samples generated at different
values of $\sin 2 \beta$ support
that the fitting procedure provides an unbiased
estimate of the value of $\sin2\beta$ of the parent distribution. 
The distribution of the difference between
the fit-$\sin2\beta$ and the true $\sin2\beta$ of the parent distribution
is well approximated by a Gaussian  and  the
fit-uncertainty on $\sin2\beta$ provides a good estimate of
the $\sigma$ of that Gaussian.

\subsection{Systematic uncertainties}

Systematic uncertainties on the measurement of $\sin 2 \beta$ due to 
flavor tagging, the
$B$ lifetime and $\Delta m_d$ are included as constraints in the fit.
We evaluated the systematic uncertainties due to  
the uncertainty in the $B^0$ mass, trigger bias and $K^0_L$ regeneration.

The systematic uncertainty arising from the $B$ mass is studied
using 1000 simulated experiments. 
The data were generated at the nominal $B$ mass and
three  full likelihood fits were  performed on each experiment.
One fit was performed 
using the normalized mass calculated with 
the nominal  $B$ mass  and two additional fits were performed
using  $B$ masses shifted by $\pm 1$ MeV/$c^2$. 
The shifts observed in $\sin 2 \beta$ 
from  fits to the simulated experiments  are consistent with  a random
distribution centered on zero with an RMS of $0.019$. 
The change in the observed RMS spread of  $\sin 2 \beta$ is $ < 0.019$
when combined in quadrature. 
We also fit  the data with the $B$ mass shifted by 1 MeV/$c^2$ and found
the value of $\sin 2 \beta$ changed by $0.013$, which consistent with
the simulation results.
We conclude the additional uncertainty  on $\sin 2 \beta$  due 
to the uncertainty on the $B$ mass 
is $ < 0.019$ and is negligible.

The data are  assumed to be  a $50$:$50$ mix of $B^0/\overline{B}^0$. 
A possible charge bias arising from the trigger is considered.
Events that are triggered on the two muons from the $J/\psi$ decay
do not contribute to the charge bias. The remaining 
30\% contain some events in which the trigger was from one of the
$J/\psi$ muons and the other lepton  candidate was from the opposite side $B$.
The magnitude of the charge bias in the trigger has been measured 
to be $< 1 \%$ at a  threshold of $P_T=2$~GeV/$c$ and is consistent
with zero for $P_T > 3$~GeV/$c$, rendering this uncertainty negligible.

Possible contamination of our
data from $K^0_L$ regeneration from the material in the
inner detector has been considered. 
Reconstruction of the $K^0_L$ as a $K^0_S$
causes the event to be entered with the incorrect sign in the asymmetry. 
This effect shifts $\sin 2 \beta$ by less than 0.003, which is
neglected.
The results of the systematic studies are shown in Table~\ref{ta:sys}.

We have evaluated the 
contribution to the sample from $B^0\rightarrow J/\psi K^*$,
with $K^*\rightarrow K^0_S \pi^0$ and the $\pi^0$ not reconstructed
and find it to be a negligible contribution.  The same is true with
$\Lambda_B\rightarrow J/\psi  \Lambda$ and $\Lambda \rightarrow p\pi^-$
and the $\Lambda$ reconstructed as $K^0_S\rightarrow \pi^+ \pi^-$;
$B_s \rightarrow J/\psi \phi$, $\phi\rightarrow K^0_S K^0_L $; and 
$B_s \rightarrow J/\psi K^0_S$.

Many checks of the data and analysis have been performed to
increase our confidence in the result. In order
to check the sensitivity of the result to the
dilutions, we  imposed alternative
JETQ and SLT dilution parameters taken from our various mixing analyses
that use the inclusive lepton sample~\cite{owen}.
We observe the 
expected shift  in the value of  $\sin 2 \beta$ and
small changes in the uncertainty.
The signal sample selection criteria have been
varied, and other than a sensitivity to the  SST
tag track $P_T$ threshold, as discussed in Ref.~\cite{ken},
we find no unexpected sensitivity in the result.

\subsection{Final result}

The maximum likelihood function fitting procedure
returns a stable value for $\sin 2 \beta$
and the  uncertainties are approximately Gaussian.
Even though asymmetric dilutions are permitted in the
fit, no significant asymmetry is observed.
Furthermore, the background asymmetries are consistent with zero.

Using the entire data set
and three tagging algorithms, we find 
\[
\sin 2\beta = 0.79 {+0.41\atop-0.44}.
\]

\begin{figure}
\postscript{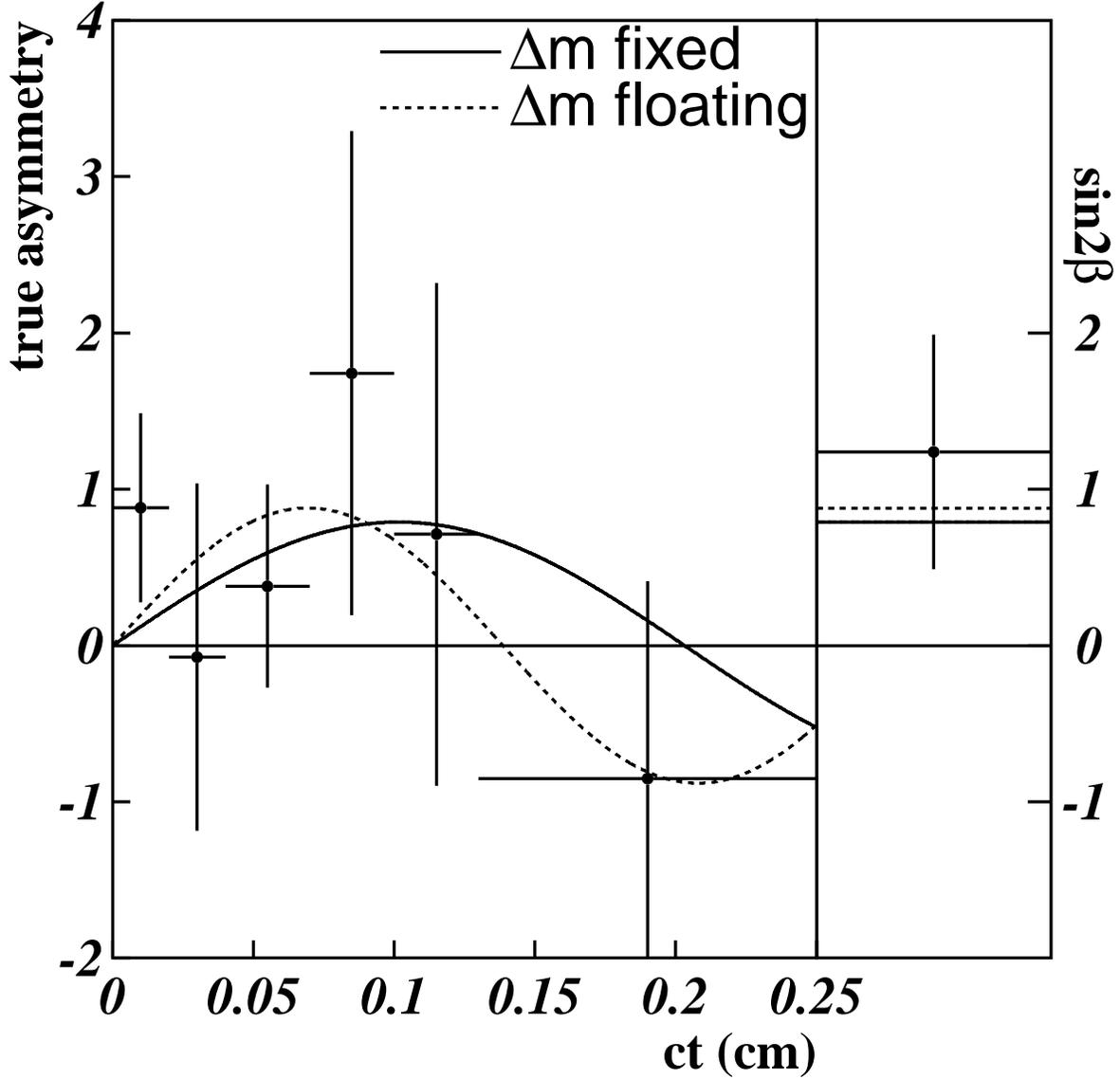}{1.0}
\caption[]{
The true asymmetry ($\sin 2 \beta \sin \Delta m_d t$) as a function of lifetime
for $B\rightarrow J/\psi K^0_S$ events.  The data points
are sideband-subtracted and have been combined according
to the effective dilution for single and double-tags.
The non-SVX events are shown on the right.}
\label{act}
\end{figure}

The asymmetry 
is shown in Fig.~\ref{act} for the SVX and non-SVX 
events separately.  The asymmetry for the SVX events
is displayed as a function of lifetime, while the asymmetry
for the non-SVX events is shown in a single, time-integrated
bin
since the decay length  information is
of low resolution. 
Although plotted as a time-integrated point, lifetime
information for the non-SVX
events is utilized in the maximum likelihood function.
The positive asymmetry preferred by the fit
can be seen. 
The curves displayed in the plot are the results from the
full maximum likelihood fit using all data. 
In order to display the data, we have
combined the effective dilution for single and double-tag events
after having subtracted the background.
The full maximum likelihood fit uses the SVX and non-SVX samples 
and treats properly the decay length, dilution 
and uncertainty for each event.

The uncertainty can be divided into statistical and systematic terms:
\[
\sin 2\beta = 0.79 \pm 0.39 (\rm stat)  \pm 0.16 ( syst).\]
The systematic term predominantly reflects the uncertainty in the result
due to the uncertainty in the dilution parameters. Although
the dilution parameters are not precisely determined, due to the
limited statistics of the $B^\pm \to J/\psi K^{\pm}$ calibration
sample, this uncertainty term  does not dominate the overall 
uncertainty on $\sin 2 \beta$. 
Furthermore, the uncertainty on 
$\sin 2 \beta$ will not be dominated by the uncertainty
on the dilution parameters in future runs because
the uncertainty scales  inversely 
with increasing statistics of the calibration
samples.

\begin{figure}
\postscript{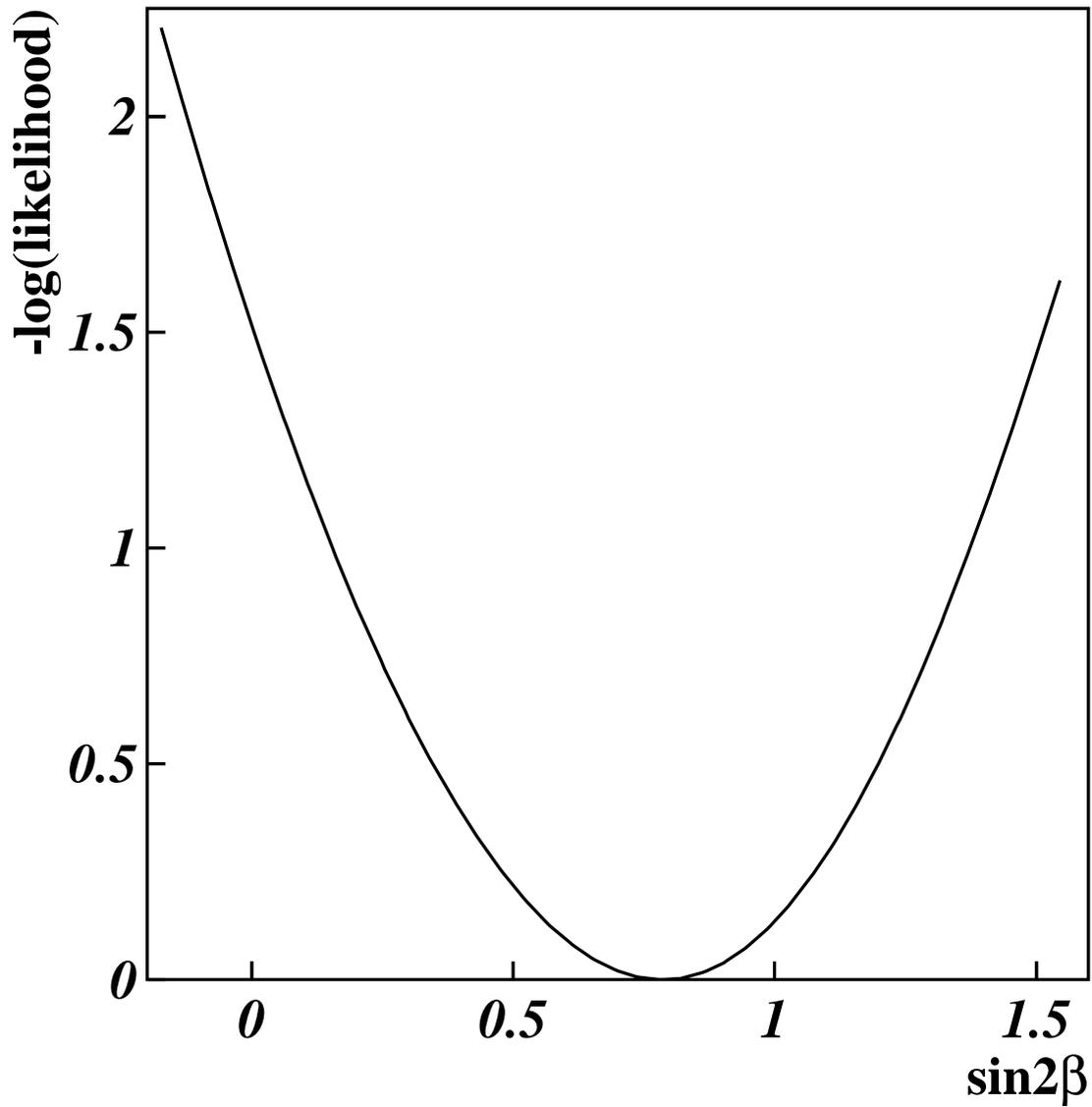}{1.0}
\caption[]{A scan of the log-likelihood function. 
The value of $\sin 2 \beta$ is scanned, and at each step,
the function is minimized.}
\label{fig:lmin}
\end{figure}

It is of interest to determine the quantitative statistical 
significance  of whether this result supports
$\sin 2 \beta > 0.0$ and hence provides an indication of
$CP$ symmetry violation in the $b$ quark system.
A scan through the likelihood
function as  $\sin 2 \beta$ is varied is shown in Fig.~\ref{fig:lmin}
and demonstrates that the uncertainties follow Gaussian statistics.
Using the Feldman-Cousins frequentist approach~\cite{feldman},
we calculate a 
confidence interval of $0.0<\sin 2 \beta < 1$  at 93\%.
An alternative approach is the  Bayesian method,  where 
a flat prior distribution in $\sin 2 \beta$ is assumed and 
a  probability that $\sin 2 \beta > 0.0$ of 95\% is calculated. 
Finally, if the true value of $\sin 2 \beta$ is  zero,  and 
the measurement uncertainty is  $0.44$ (Gaussian uncertainty), the 
probability of obtaining $\sin 2 \beta > 0.79$ is 3.6\%. 
This value is obtained by simply integrating the Gaussian
distribution from $0.79$ to $\infty$. 
The toy Monte Carlo is in good agreement with the calculated probability.

\begin{figure}
\postscript{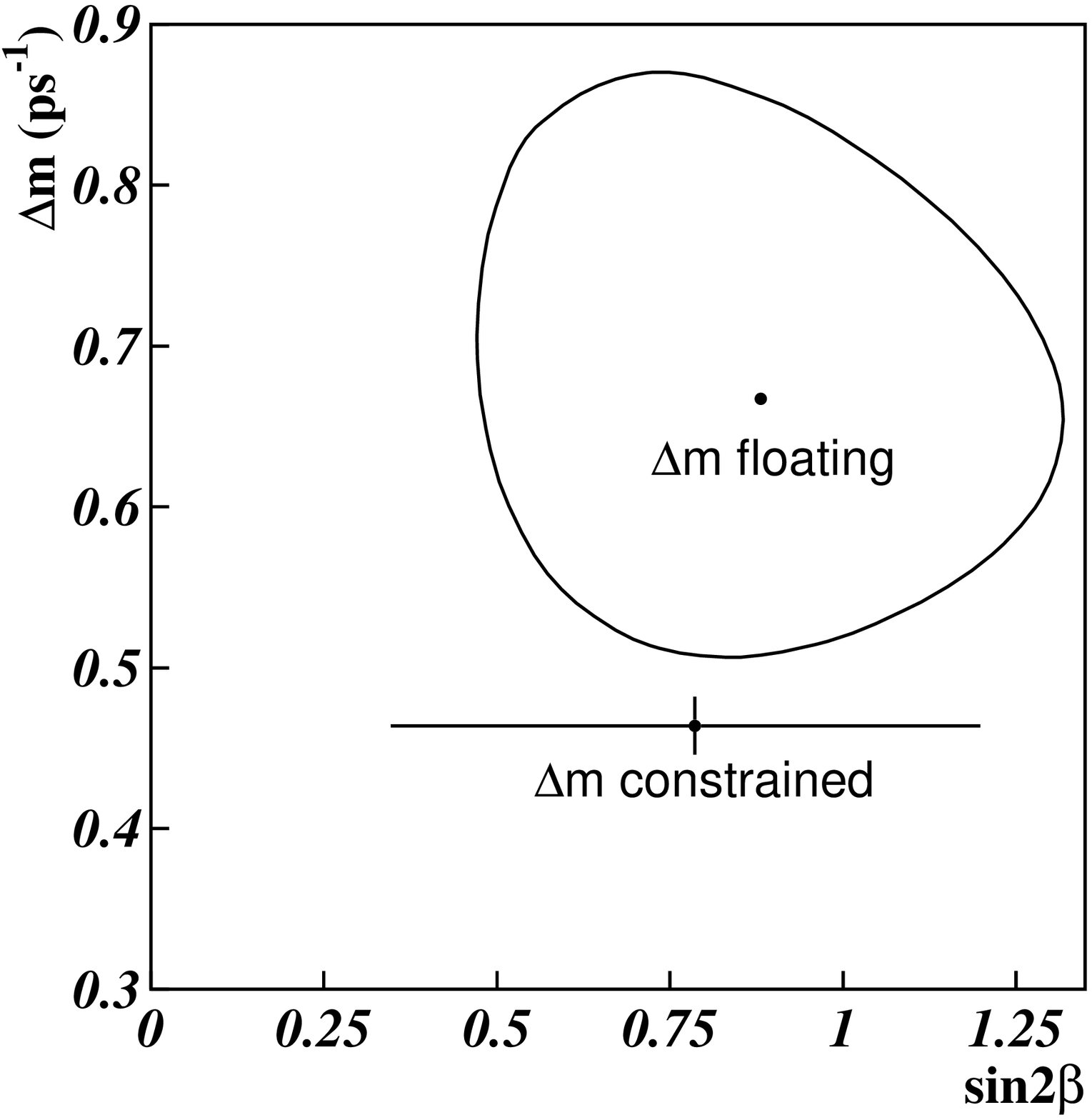}{1.0}
\caption[]{
The $1\sigma$ (39\%) $\sintbetasym$-$\deltambosym$ contour from
a fit with $\deltambosym$ constrained only by the $B \to J/\psi K^0_S$ data.
Also shown is the nominal fit with  
$\Delta m_d = (0.464\pm 0.018)\, \hbar \rm ps^{-1}$~\cite{pdg}.} 
\label{floatdm}
\end{figure}

It is possible to remove the constraint that ties $\Delta m_d$ to
the world average 
value  and to fit
for $\sintbetasym$ and $\deltambosym$ simultaneously. In this case
the result is $\sin 2 \beta = 0.88 {+0.44\atop-0.41}$ and
$\Delta m_d=0.68\pm0.17\, \hbar \rm ps^{-1}$.
The value of $\Delta m_d$ from the fit agrees with the world value 
at the level of $\sim \! 1.2\sigma$. 
This agreement increases our confidence in the main result.
Figure~\ref{floatdm} shows the $1\sigma$ ``error ellipse'' contour
in $\sintbetasym$-$\deltambosym$ parameter space for the fit when both
parameters float freely, and for comparison
the nominal $\sintbetasym$ result with the world average 
$\deltambosym$  and 
uncertainty.
From the roughly circular shape of the contour,
the $\deltambosym$ and $\sintbetasym$ parameters are
largely uncorrelated in the fit.

A time-integrated measurement to
check  the final result was performed.
This simplified analysis does not use the time dependence
of the asymmetry and ignores the small tagging asymmetry 
corrections applied in the
full maximum likelihood fit.
Each event falls into one of 12 classifications 
depending upon the type of flavor tags available 
for that event. Each event can be 
associated with only one class of tag combination. 
The effective tagging efficiency for the entire sample,
$\epsilon D^2$, is $(6.3\pm1.7)$\%.
A value of $\sin 2 \beta$ for each class is calculated and
a weighted average from the 12 classes is determined. Ignoring correlations
in the dilution,  $\sin 2 \beta = 0.71 \pm 0.63$.
This value is consistent with the final result and 
demonstrates the improvement in the uncertainty of $\sin 2 \beta$
provided by the full maximum likelihood procedure. This improvement
agrees well with improvements observed using the toy Monte Carlo.

Table~\ref{ta:s2b_results} summarizes fit results for
various tag-dataset combinations. 
The three tagging algorithms
contribute roughly equally to the precision of the $\sin 2 \beta$ measurement.
Although the SVX and non-SVX sample sizes are approximately equal,
the SVX events contribute more significantly to the final result.
The main reasons for this are that
the precision lifetime information from the SVX allows a better determination
of where the decay takes place along the oscillation curve 
and the better signal-to-background level from eliminating
the prompt background.

The row in Table~\ref{ta:s2b_results} 
labelled SVX SST is the result obtained when
this analysis  restricts the data set to the SVX sample and uses
only the SST algorithm.
This procedure 
essentially repeats the published CDF $\sin 2 \beta$ analysis
that obtained
$\sin 2 \beta=1.8\pm1.1(\rm stat)\pm0.3(syst)$. 
The small difference is due to sample selection.

\section{Mixing in the {\boldmath $B \to J/\psi K^*$} Sample as a Check}

A control sample
of  $B^0\rightarrow J/\psi K^*(892)^0$ decays, where
$K^*(892)^0\rightarrow K^{\pm}\pi^{\mp}$,  can be 
analyzed  for the presence of an oscillation due to
mixing ($\Delta m_d$ is well measured)
in order to verify the tag algorithms and likelihood fitting procedure.
The three flavor tagging
algorithms are used  to determine the neutral $B$ flavor at
the time of production and the dilution parameters are constrained
using the same values as in the $B \to J/\psi K^0_S$ analysis.
The charge of the  kaon is used to differentiate 
the $B^0$ from $\overline{B}^0$ at the time of decay.
After correcting for tagging dilutions, 
the amplitude of the oscillation still 
differs from unity due to the  probability
that the $K^+\pi^-$ is reconstructed as $K^-\pi^+$, which
occurs about ${P}_K=5\% $ of the time due to the 
wide $K^*$ resonance.

The  $J/\psi K_S^0$-$J/\psi K^*(892)^0$ analogy
is however not perfect. In order to achieve 
similar signal-to-background ratios, the selection criteria for the
$B \to J/\psi K^*(892)^0$ are more severe, which changes the
kinematic properties of one sample with respect to the other.
The largest backgrounds for both decay modes 
are at short decay distances and they decrease as
the flight path increases. This  works to our advantage in
the $CP$ analysis but  reduces the sensitivity
of the mixing analysis. In particular, 
due to the different oscillation phase in the $CP$ analysis versus
this  mixing analysis, 
($\sin({\Delta m_d t})\to\cos({\Delta m_d t})$), the
smallest  signal-to-background ratio  occurs at
the peak of the mixing amplitude
for $B \to J/\psi K^*(892)^0$ data set, 
where as a very favorable  signal-to-background ratio 
occurs at the peak of the $B \to J/\psi K_S^0$ oscillation.
In both the $J/\psi K_S^0$ and $J/\psi K^*(892)^0$ modes,
 $75$-$80\% $  of the background is prompt, \it i.e. \rm 
consistent with having zero lifetime.

The sample is constructed using similar criteria
to that used to reconstruct the $B \to J/\psi K$ decay modes
in this paper. 
The $J/\psi$ selection for this decay mode 
is the same as  the  $J/\psi K^0_S$ analysis.   Pion and kaon
tracks are required to have 
$P_T > 500$ MeV/$c$.  The reconstructed
$K^*(892)^0$ candidates are required to have an invariant  mass
within 80 MeV/$c^2$ of the world average of 
$896.10\pm 0.28\, {\rm MeV}/c^2$~\cite{pdg}
$K^*(892)^0$ mass. The $K^*$ candidate must have 
$P_T > 3$~GeV/$c$.   
The four-track fit for $J/\psi K^*$ is the  same as the fit for 
$J/\psi K_S^0$, except the four tracks are required to meet at
a common point and the
$K^*$ mass is not constrained.
If a candidate event has
two tracks that  satisfy two $K^*(892)^0$
combinations ($K^+\pi^-$/$K^-\pi^+$) then the 
combination with a $K\pi$ mass closest to the mean
$K^*(892)^0$ mass is chosen.  Finally, if multiple $K^*$
candidates are found in an event, the $K^*(892)^0$ candidate
chosen is the one that gives the best four-track fit.
All four charged tracks  ($\mu$,$\mu$,$K$,$\pi$)
must originate from a common vertex and  
a  $P_T(B) > 4.5$~GeV/$c$ is required.
A total signal sample of $226\pm 24$
events where both muon candidates 
have precision lifetime information and $231 \pm 28$ 
events where $\leq 1$ muon candidate  has precision
lifetime information are found.

The maximum likelihood 
fit to the $J/\psi K^*(892)^0$ data is implemented in the same way
as previously described for $J/\psi K_S^0$ except for the time-function 
$T_{\rm S}$
in which $h(t)$ is replaced by:
\[
h(t)={e^{-t/\tau}\over\tau}\epsilon_{q_1q_2}(
1+{\cal D}_K{\cal D}_{q_1q_2}\cos({\Delta m_d t})).
\]

Here ${\cal D}_K=q_KD_K$, where
$q_K$ is the charge of the $K^\pm$ from the decay of the $K^*(892)^0$,
and $D_K$ is the dilution arising from the inability
to correctly distinguish the charged kaon from the charged pion
in the $K^*(892)^0$ decay. 
The dilution $D_K$ is the free parameter 
in this fit and 
is analogous to $\sin2\beta$ in the $J/\psi K_S^0$ fit, the parameters
in each case representing the amplitude of an oscillation.
The amplitude is expected to be
$D_{K} = 1-2P_{K}= 0.9{+0.1\atop-0.2}$ where $D_{K}$ 
is the dilution factor coming from incorrect $K$-$\pi$
assignment~\cite{petar}.

\begin{figure}
\postscript{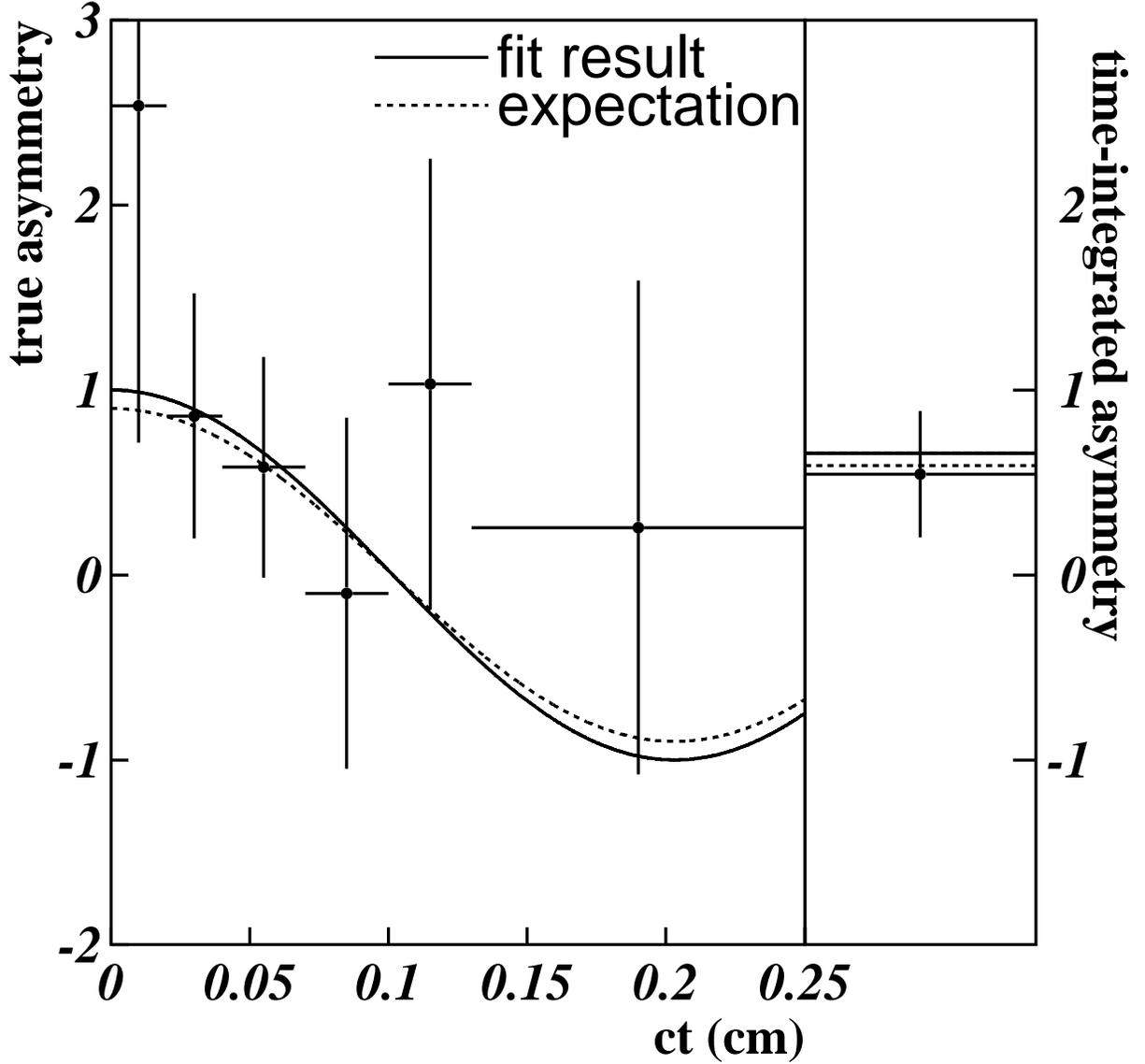}{1.0}
\caption[]{The true asymmetry 
($D_K \cos{\Delta m_d t}$)
as a function of lifetime
for  $B^0\rightarrow J/\psi K^*(892)^0$ events.  The data points
are sideband-subtracted and have been combined according
to the effective dilution for single and double-tags.
The time-integrated asymmetry for non-SVX events is shown on the right. 
The solid curve represents the 
maximum likelihood fit in which $\Delta m_d$ is fixed and
the dashed curve is the expectation when we also
fix $D_K$. }
\label{kstar}
\end{figure}

When $\Delta m_d$ is fixed to the world average, we measure
$D_{K} = 1.00 \pm 0.37$, which is consistent with 
expectation. When $\Delta m_d$ is allowed to
float, we measure: $D_{K} = 0.96 \pm 0.38$ and
$\Delta m_d = 0.40 \pm 0.18 \hbar$ ps$^{-1}$, which is consistent
with the world average
$\Delta m_d = (0.464\pm 0.018)\, \hbar \rm ps^{-1}$~\cite{pdg}. 
The results of the fits 
are shown in Fig~\ref{kstar}.

Although the statistics are
not sufficient for a precise measurement of $\Delta m_d$, 
this check on an independent sample of events is entirely
consistent with our expectation.

\section{Conclusion}

We have presented a  measurement of $\sin 2 \beta$ using
$\sim \! 400$ $B  \rightarrow J/\psi K^0_S$ 
events reconstructed with the CDF detector.  We find:
\[
\sin 2\beta = 0.79 {+0.41\atop-0.44}\rm (stat+syst)
\]
with the uncertainty dominated by the statistical contribution.

We have calculated
the statistical significance  of 
whether this result supports
$\sin 2 \beta > 0.0$ and hence provides indication for
$CP$ symmetry violation in the $b$ quark system.
Using the Feldman-Cousins~\cite{feldman} method,
a 93\% confidence interval of
$0.0<\sin 2 \beta <1.00$ is found.  Alternative methods
yield similar limits.
This  measurement  is the best direct indication  that  $CP$ invariance
is violated in the $b$ quark system 
and is consistent with the
Standard Model expectation of a large 
positive value of  $\sin 2 \beta$~\cite{nierste,london,delphi,mele}.
The sign of our result
supports the favored positive signs for $B_B$ and $B_K$. 
With an anticipated luminosity of $2\, \rm fb^{-1}$ in 
Run~II, we expect, based on a simple extrapolation of this measurement,
an uncertainty on $\sin 2 \beta$ of $\sim \! 0.08$.
Detector upgrades in progress should further reduce this
uncertainty.

\acknowledgments

     We thank the Fermilab staff and the technical staffs of the
participating institutions for their vital contributions.  This work was
supported by the U.S. Department of Energy and National Science Foundation;
the Italian Istituto Nazionale di Fisica Nucleare; the Ministry of Education,
Science and Culture of Japan; the Natural Sciences and Engineering Research
Council of Canada; the National Science Council of the Republic of China; 
the Swiss National Science Foundation; the A. P. Sloan Foundation; and the
Bundesministerium f\"{u}r Bildung und Forschung, Germany.



\begin{table}[bthp]
\squeezetable
\caption{Summary  of tagging algorithms performance.
All numbers listed are in percent. 
The efficiencies  are obtained from the $B \to J/\psi K^0_S$ sample.
The dilution information
is derived from the $B^\pm \to J/\psi K^{\pm}$ sample.
}
\label{ta:taggers}
\begin{tabular}{ccccc}
tag side &tag type& class & efficiency & dilution  \\
\hline 
same-side &
SST
& $\mu_1,\mu_2$ in SVX & $35.5\pm 3.7$ &
$16.6\pm2.2$  \\
& 
SST
& $\mu_1$ or $\mu_2$ non-SVX & $38.1 \pm 3.9$ &
$17.4 \pm 3.6$  \\  
opposite side &
SLT
& all events & $5.6\pm 1.8$ &
$62.5 \pm 14.6$  \\
& 
JETQ
& all events &$40.2 \pm 3.9$ & 
 $23.5 \pm 6.9$  \\
\end{tabular}
\end{table}

\begin{table}[hbtp]
\caption{Definition of tags.  For the case of the SST algorithm,
the tag depends upon the charge of a track ($t^+$,$t^-$)
near the $B$; for the SLT algorithm, the tag depends upon the
charge of a lepton in the event ($\ell^+$,$\ell^-$); for
the JETQ algorithm, the tag depends upon the
average weighted charge  of tracks in 
a jet ($Q_{\rm jet}$).}
\label{ta:tagdefs}
\begin{tabular}{cccc}
tag & positive ($+$) tag  & negative ($-$) tag  & no tag \\
 & $B^0\rightarrow J/\psi K^0_S$ & $\overline{B}^0 \rightarrow 
J/\psi K^0_S$ & \\
 \hline
SST & single track $t^+$ & single track $t^-$ & no track \\
SLT  & single lepton $\ell^-$ & single lepton $\ell^+$ & no lepton \\
JETQ  & $Q_{\rm jet}<-0.20$ & $Q_{\rm jet}>0.20$ &
   $ |Q_{\rm jet}| \leq 0.20$ \\
\end{tabular}
\end{table}

\begin{table}[ht]
\squeezetable
\caption{The dilutions determined from the $B^\pm \to J/\psi K^{\pm}$
sample and the efficiency ratios determined from the
inclusive $J/\psi$ sample are shown.  $D_{\rm ave}$ is the average dilution.
The SST dilutions utilize additional information as described
in the text.}
\label{ta:sumall}
\begin{tabular}{lcccc}
tag & $\epsilon_+/\epsilon_-$ & $D_+ (\%)$ & $D_-(\%)$ & $D_{\rm ave}(\%) $ \\
\hline
$\rm SST_{SVX}$ & $1.031 \pm 0.011$ & $16.1 \pm 5.1$ & $ 17.1 \pm 5.2 $ & 
$16.6\pm 2.2$ \\
$\rm SST_{\text{non-SVX}}$ & $1.037 \pm 0.010$ &
    $17.0\pm 5.7$ & $ 17.8 \pm 5.8 $ &
$17.4\pm 3.6$ \\
SLT  & $0.978 \pm 0.047$ & $76.9 \pm 19.6$ & $ 46.4 \pm 21.8 $ & 
$62.5\pm 14.6$ \\
JETQ  & $0.977 \pm 0.015$ & $20.7 \pm 9.3$ & $ 26.5 \pm 8.3 $ & 
$23.5\pm 6.9$  \\
\end{tabular}
\end{table}

\begin{table}[ht]
\caption{
Systematic uncertainties in the measurement of $\sin 2 \beta$.
The items labelled ``in fit'' are parameters that are allowed
to float in the fit but are constrained by their measured uncertainties.
The uncertainty returned from the likelihood fit includes the contributions
from these sources.\label{ta:sys}}
\begin{tabular}{ccc}
parameter  & $\delta \sin2\beta$ & in fit\\
\hline
dilution and efficiency & 0.16 & yes \\
$\Delta m_d$            & negligible & yes \\
$\tau_{B^0}$            & negligible & yes \\
$m_B$                   & negligible & yes  \\
trigger bias            & negligible & no \\
$K^0_L$ regeneration    &  negligible & no\\
\end{tabular}
\end{table}

\begin{table}[h]
\caption{
Fit $\sin2\beta$ results for the three  tagging algorithms. The
combined $\chi^2$ for the SST, JETQ, and SLT tagging
algorithms  is 4.63 for
2 degrees of freedom, giving a probability of $\sim \! 10$\%.
\label{ta:s2b_results}}
\begin{tabular}{lrrrr}
data & tag(s) &  $\sin2\beta$ & $+$ error & $-$ error\\
\hline
all & all & 0.79 & 0.41 & 0.44 \\
    & SST & 2.03 & 0.84 & 0.77 \\
    & JETQ & $-$0.31 & 0.81 & 0.85 \\
    & SLT & 0.52 & 0.61 & 0.75 \\
SVX & all & 0.54 & 0.52 & 0.57 \\
    & SST & 1.77 & 1.04 & 1.01 \\
non-SVX & all & 1.24 & 0.75 & 0.70 \\
\end{tabular}
\end{table}


\begin{references}

\bibitem{cronin} J.H.~Christenson {\it et al.},
Phys.\ Rev.\ Lett.\ {\bf 13}, 138 (1964).

\bibitem{sanda} A.B.~Carter and A.I.~Sanda,
Phys.\ Rev.\ Lett.\ {\bf 45}, 952 (1980);
Phys.\ Rev.\ D {\bf 23}, 1567 (1981).

\bibitem{bigi} I.I.~Bigi and A.I.~Sanda,
Nucl.\ Phys.\ {\bf B193}, 85 (1981);
Nucl.\ Phys.\ {\bf B281}, 41 (1987).

\bibitem{Khoze} I.I.~Bigi, V.A.~Khoze, N.G.~Uraltsev, and A.I.~Sanda,
in $CP$ {\it Violation}, edited by C.~Jarlskog (World Scientific,
Singapore, 1989), p.~175.

\bibitem{rosnerisi} I.~Dunietz and J.L.~Rosner, 
Phys.\ Rev.\ D {\bf 34}, 1404 (1986).

\bibitem{quinn} For a review of $CP$ violation in $B$ decays,
Y.~Nir and H.R.~Quinn, Annu.\ Rev.\ Nucl.\ Part.\ Sci.\ {\bf 42}, 221 (1992).

\bibitem{despande} N.G.~Deshpande, X.G.~He, and S.~Oh,
Z.\ Phys.\ C {\bf 74}, 359 (1997).

\bibitem{pdg}
Particle Data Group, C.~Caso {\it et al.},
Eur.\ Phys.\ J.\ C {\bf 3}, 1 (1998).

\bibitem{opal} K.~Ackerstaff {\it et al.},
Eur.\ Phys.\ J.\ C {\bf 5}, 379 (1998).

\bibitem{ken} CDF Collaboration, F.~Abe {\it et al.},
Phys.\ Rev.\ Lett.\ {\bf 81}, 5513 (1998);
K.~Kelley, Ph.D. Dissertation,
Massachusetts Institute of Technology, 1999 (unpublished).

\bibitem{ckm} N.~Cabibbo,
Phys.\ Rev.\ Lett.\ {\bf 10}, 531 (1963);
M.~Kobayashi and T.~Maskawa,
Prog.\ Theor.\ Phys.\ {\bf 49}, 652 (1973).

\bibitem{lincoln} L.~Wolfenstein,
Phys.\ Rev.\ Lett.\ {\bf 51}, 1945 (1983).

\bibitem{linglee} L.-L.~Chau and W.-Y.~Keung,
Phys.\ Rev.\ Lett.\ {\bf 53}, 1802 (1984); 
C.~Jarlskog and R.~Stora, Phys.\ Lett.\ B {\bf 208}, 268 (1988);
J.D.~Bjorken, Phys.\ Rev.\ D {\bf 39}, 1396 (1989).

\bibitem{nierste} S.~Herrlich and U.~Nierste,  
Phys.\ Rev.\ D {\bf 52}, 6505 (1995).

\bibitem{london} A.~Ali and D.~London,
Nucl.\ Phys.\ B (Proc. Suppl.) {\bf 54A}, 297 (1997).

\bibitem{delphi} P.~Paganini {\it et al.},
Phys.\ Scr.\ {\bf 58}, 556 (1998). 

\bibitem{mele} S.~Mele,
Phys.\ Rev.\ D {\bf 59}, 113011 (1999).

\bibitem{kayser} Y.~Grossman, B.~Kayser, and Y.~Nir,
Phys.\ Lett.\ B {\bf 415}, 90 (1997);
I.I.~Bigi and A.I.~Sanda,
Phys.\ Rev.\ D {\bf 60}, 033001 (1999.).

\bibitem{detector} CDF Collaboration, F.~Abe {\it et al.},
Nucl.\ Instrum.\ Methods Phys.\ Res.\ A {\bf 271}, 387 (1988).

\bibitem{TopPRD} CDF Collaboration, F.~Abe {\it et al.},
Phys.\ Rev.\ Lett.\ {\bf 74}, 2626 (1995);
see also Phys.\ Rev.\ D {\bf 50}, 2966 (1994).

\bibitem{SVX} D.~Amidei {\it et al.},
Nucl.\ Instrum.\ Methods Phys.\ Res.\ A {\bf 350}, 73 (1994); 
P.~Azzi {\it et al.},
{\it ibid.} {\bf 360}, 137 (1995).

\bibitem{julio} CDF Collaboration, F.~Abe {\it et al.},
Phys.\ Rev.\ Lett.\ {\bf 76}, 2015 (1996).

\bibitem{pekka} CDF Collaboration, F.~Abe {\it et al.},
Phys.\ Rev.\ D {\bf 54}, 6596 (1996).

\bibitem{hopkins} CDF Collaboration, F.~Abe {\it et al.},
Phys.\ Rev.\ D {\bf 57}, 5382 (1998).

\bibitem{petar} CDF Collaboration, F.~Abe  {\it et al.},
Phys.\ Rev.\ Lett.\ {\bf 80}, 2057 (1998);
Phys.\ Rev.\ D {\bf 59}, 032001 (1999);
P.~Maksimovi\'{c}, Ph.D. Dissertation,
Massachusetts Institute of Technology, 1998 (unpublished).

\bibitem{rosner} M.~Gronau, A.~Nippe, and J.L.~Rosner,
Phys.\ Rev.\ D {\bf 47}, 1988 (1993);
M.~Gronau and J.L.~Rosner,
{\it ibid.} {\bf 49}, 254 (1994).

\bibitem{owen} CDF Collaboration, F.~Abe {\it et al.},
Phys.\ Rev.\ D {\bf 60}, 072003 (1999);
O.~Long, Ph.D. Dissertation, University of 
Pennsylvania, 1998 (unpublished);
M.~Peters, Ph.D. Dissertation, University
of California, Berkeley, 1998 (unpublished).

\bibitem{JADE} W.~Bartel {\it et al.},
Z.\ Phys.\ C {\bf 33}, 23 (1986).


\bibitem{MINUIT}
F.~James, {\it MINUIT---Function Minimization and Error Analysis},
Version 94.1, CERN Program Library Long Writeup D506,
CERN, Geneva, Switzerland, 1994.

\bibitem{feldman} G.J.~Feldman and R.D.~Cousins,
Phys.\ Rev.\ D {\bf 57}, 3873 (1998).

\end{references}
\end{document}